\documentclass{article}

\usepackage{arxiv}

\usepackage[utf8]{inputenc} % allow utf-8 input
\usepackage[T1]{fontenc}    % use 8-bit T1 fonts
\usepackage{hyperref}       % hyperlinks
\usepackage{url}            % simple URL typesetting
\usepackage{booktabs}       % professional-quality tables
\usepackage{amsfonts}       % blackboard math symbols
\usepackage{nicefrac}       % compact symbols for 1/2, etc.
\usepackage{microtype}      % microtypography
\usepackage{lipsum}		% Can be removed after putting your text content
\usepackage{graphicx}
\usepackage{amsmath}
\usepackage{float}
\usepackage{subcaption}
\title{Simulating the influence of sea-surface-temperature (SST) on tropical cyclones over South-West Indian ocean, using the UEMS-WRF regional climate model}

\author{
  Chibueze N. Oguejiofor\thanks{Use footnote for providing further
    information about author (webpage, alternative
    address)---\emph{not} for acknowledging funding agencies.} \\
  African Institute for Mathematical Sciences\\
  AIMS, Rwanda.\\
  Rwanda, PA 15213 \\
  \texttt{oguejiofor.chibueze@aims.ac.rw} \\
   \And
 Babatunde J.~Abiodun \\
 University of Capetown\\
  Capetown, South Africa.\\
  South Africa,7700 \\
  \texttt{babiodun@csag.uct.ac.za} \\
}

\begin{document}
\maketitle

\begin{abstract}
Tropical cyclones remain a major threat to the lives, property and economy of communities around the South West Indian ocean, notably Southern Africa and Madagascar. Tropical cyclones occur quite frequently around this region, hence the need to be adequately prepared for the impact of climate variability and climate change on them. Previous studies have shown that global warming due to anthropogenic forces may lead to an increase in the frequency of occurrence of tropical cyclones within the region. However, there is little work done regarding the impact of this on the intensity of tropical cyclones formed. This study uses the weather research forecast (WRF) model to perform a series of simulations for tropical cyclone Enawo with the aim of investigating the effect of an increase or decrease (by 2$^{\circ}C$) in sea surface temperature (SST) on the intensity of the tropical cyclone (using windspeed, precipitation and pressure as measures of cyclone intensity). The experiment uses the data from European Centre for Medium-Range Weather Forecasts (ECMWF) ERA5 re-analysis dataset to validate the results of the WRF model which was ran using boundary conditions data from climate forecast system reanalysis  (CFSR).

The results indicate that the WRF model performs reasonably well in simulating the track and windspeed of the tropical cyclone, when compared to observational data. In simulating the tropical cyclone, the WRF model shows that an increase in the SST by 2$^{\circ}C$ generally increases the intensity of the tropical cyclone formed. This is evident in the increasing maximum precipitation rate as well as windspeed, and decreasing minimum pressure. An increase in SST also causes the emergence of a second low pressure system. On the other hand, a decrease in the SST by 2$^{\circ}C$ leads to a minute effect in the intensity but generally acts to decrease it. This results in a smoother track path for the tropical cyclone, a decrease in the maximum precipitation rate and windspeed, and an increase in the surface pressure.

The results of this study have shown that increasing the global temperature by around 2$^{\circ}C$ - violation of the \textit{Paris Accord} - would lead to more violent and unpredictable tropical cyclones within the South West Indian ocean, and hence more destruction and loss of lives.
\end{abstract}

% keywords can be removed

\section{Introduction}
\subsection{Background}

Tropical cyclones are one of the most catastrophic natural phenomena in occurence today. They are a deadly force of nature, wrecking havoc to coastal cities as well as inland. The ripple effect of tropical cyclone activities range from severe and destructive wind, flooding, landslides/mudslides, pollution of fresh water with sewage, storm surges, enhancing mosquito-borne diseases, amongst others.
Tropical cyclones account for nine of the ten most costly inflation-adjusted insurance natural disaster losses (200m dollars) between 1970 and 2009 \cite{2010weathering}. 
This is is because strong windspeed from tropical cyclones have been known to destroy overhead powerlines (leading to electrocution), drown vehicles and complete overrun of low-rise building structure. Of particular concern, is the effect of tropical cyclones on tall buildings/skyscrapers in which it has been well known to distort the foundation of skyscrapers, hence, leading to a tilt in buildings structures and a domino effect of a series of skyscrapers on one another.
In original loss values, tropical cyclones account for two of the five most costly economic losses and four of the five most costly insurance losses from natural disasters over the period 1950 to 2009 \cite{2010natural}.

Apart from  the loss of property, the loss of lives due to tropical cyclones have been increasingly alarming over the last decade. A total of about 10,000 people have died each year as a direct result of tropical cyclones or other destructions emanating from it \cite{adler2005estimating}.
The deadliest tropical cyclone ever recorded was the cyclone Bhola which occurred in the 1970. This cyclone had an estimated death toll of between  300,000 and 500,000 lives.
The most recent of such destructive nature of tropical cyclones occurred between 4th to 21st  of March 2019 as a tropical cyclone named Idai made landfall twice and affected a total of 4 countries (Madagascar, Mozambique, Malawi and Zimbabwe) with a death toll of over 1000 people and several thousands more still missing. This was one of the most deadliest cyclones in the South-West Indian ocean with a damage of  >1billion US dollars \cite{crash}.
Countries around the South-West Indian ocean (notably Madagascar and Mozambique) are particularly vulnerable to cyclones. One of the most critical factors needed for the formation of tropical cyclones is a warm sea temperature. Hence, there is need for a proper understanding of the effect of climate change (notably sea-surface-temperature increase) on tropical cyclogenesis and variation of it’s intensity and frequency with increasing sea surface temperature.

The intensity and frequency of natural disasters have generally  increased in recent decades more than humans are able to adjust to. This has largely been associated with anthropogenic forces. Natural disasters possess an inherent unpredictability that makes preparation and planning an elusive task for countries. One of these natural disasters are Tropical cyclones also known as Hurricanes, Typhoons or Tornado’s. The choice of name is dependent on the location (North-West Pacific – Typhoon, Atlantic and North-East Pacific – Hurricanes, South Pacific – Cyclones).

\begin{figure}[!h]
% Use "\centering" in floats (figure, table), but if you need to center
% some text (why?) use "\begin{center}...\end{center}".
\centering 
% Figure environments same as 0.8 * \textwidth please
% That does not necessarily mean the actual picture size,
% it is a guideline for the environment which could contain
% 2 or more pictures! Be consistent and follow the guidelines
% provided in your sources.
\includegraphics[width=0.8\textwidth]{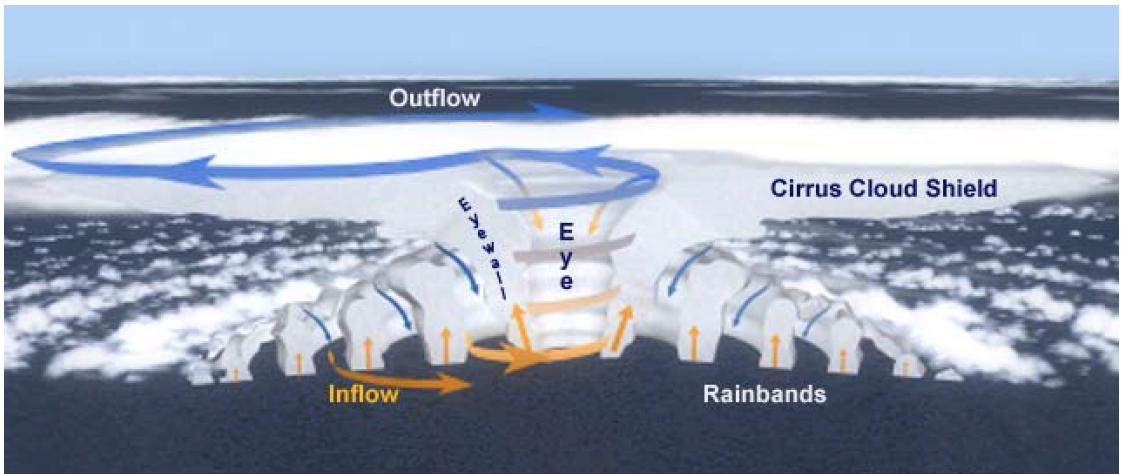}
\caption{Cross section of the structure of a cyclone showing the flow of air and the location of the cyclone eye and eye wall}
\label{chi} 
% if you move the label it breaks the reference numbering; 
% always have it *after* the caption.
\end{figure}

Tropical cyclones are basically strong spiraling mass of wind about a low pressure core in the South Pacific (Figure~\ref{chi}). They have been known to cost billions of dollars in damages around various regions around the world. They exist as a warm-core, self-sustaining spiral of wind with surface wind reaching up to 110$mph$. Tropical cyclones usually originate in warm sea waters (temperature >26.5$^{\circ}C$) before moving and generating strong winds which are usually most destructive when they make landfall. 

Tropical cyclones are categorized on a scale of 1 to 5 based on the amount of damage caused due to wind gush and speed - category 5 being the most destructive with wind gusts greater than 280$km h^{-1}$ and category 1 being the least destructive with wind gusts between 90-125 $km h^{-1}$. 
Apart from classification based on destruction, cyclones are classified into stages characterized by wind force. These include tropical disturbance, tropical depression, moderate tropical storm, severe tropical storm, tropical cyclone, intense tropical cyclone and very intense tropical cyclone with wind-force of < 28$kt$ (<5 $km h^{-1}$), 28-33$kt$ (51-63 $km h^{-1}$), 34-47$kt$ (63-88 $km h^{-1}$), 48-63$kt$ (89-117 $km h^{-1}$), 64-89$kt$ (118-165 $km h^{-1}$), 90-115$kt$ (166-212 $km h^{-1}$) and >115$kt$ (>212 $km h^{-1}$) respectively.

\subsection{Statement of problem}
The effect of anthropogenic forces in altering the climate has been profound, warranting the \textit{Paris accord} in 2016. This agreement signed by over 196 state parties had a goal to keep the increase in global average temperature to well below 2$^{\circ}C$; and to limit the increase to 1.5$^{\circ}C$.

The result of sea surface temperature (SST) variation on the frequency of cyclone is not as consistent as it’s effect on the cyclone intensity \cite{knutson2010tropical}.
Observational data shows that the ideal temperature for the formation of cyclones is >26.5$^{\circ}$C, hence, the need to investigate how increasing/decreasing SST could affect the intensity of cyclones formed, especially in the South Western Indian ocean. The parameters used in the quantification of cyclone intensity includes rainfall, pressure, wind and precipitation rate.

\subsection{Aim and objectives}
The aim of this research is to understand and compare the effect of increasing/decreasing sea surface temperature (SST) on the intensity of tropical cyclones formed around the South-West Indian ocean using the WRF model. This will help identify our expectation on the effect of global warming around this region, if the Paris agreement fails to be reached.

The objectives of the study is:
\begin{itemize}
\item Identify the rainfall, pressure and wind pattern of cyclone Enawo just as it was about to make landfall in Madagascar.
\item Simulate the influence of increased temperature on the rainfall, pressure andwind speed on cyclone Enawo.
\item Simulate the influence of decreased temperature on the rainfall, pressure and wind speed on cyclone Enawo.
\item Make a detailed comparison on the effect of sea surface temperature change on the intensity of tropical cyclones formed.

\end{itemize}

\subsection{Description of study area}
\subsubsection{Location of study area}
\begin{figure}[!h]
\centering 
\includegraphics[width=0.8\textwidth]{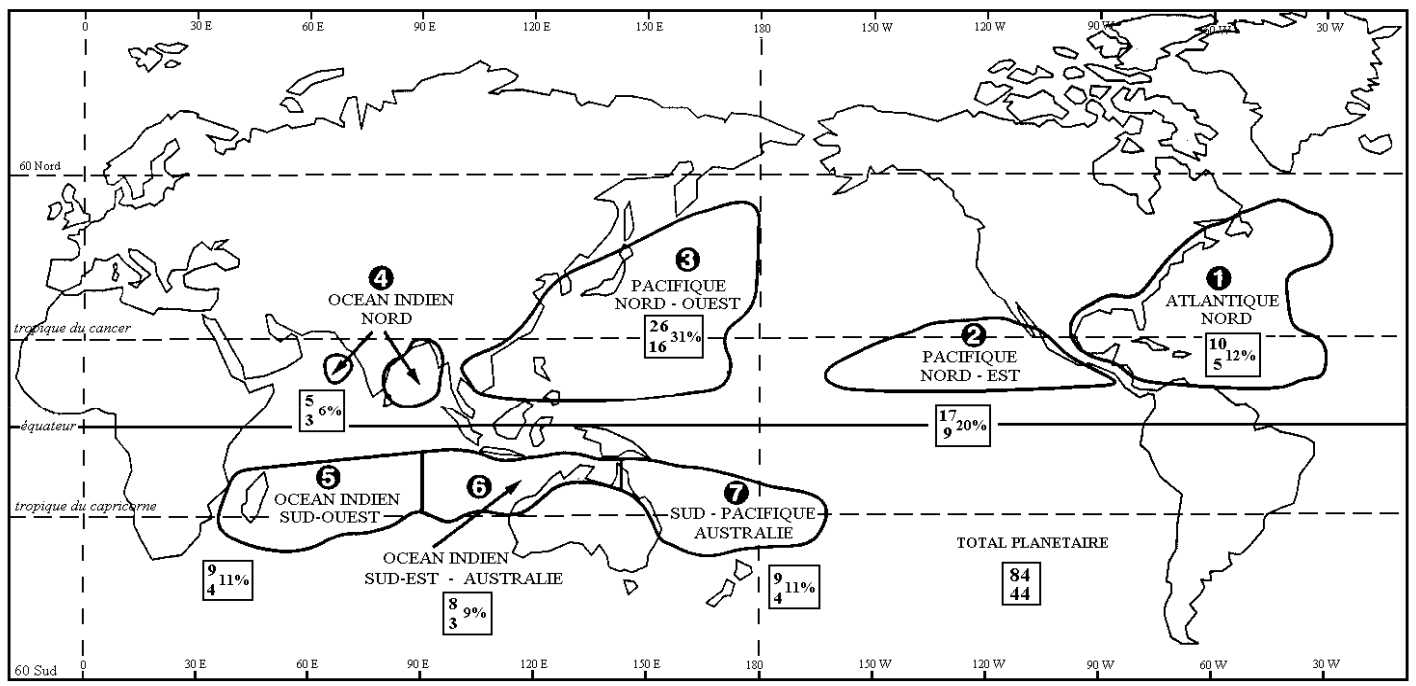}
\caption{Global distribution of cyclone basin in the world and the statistics of average annual number of tropical storms and cyclones, average annual number of tropical cyclones and percentage of the population recorded between 1968 and 1990 \cite{neumann1993global}.}
\label{1} 
% if you move the label it breaks the reference numbering; 
% always have it *after* the caption.
\end{figure}

The South-West Indian basin is one of seven known tropical cyclone basins in the world today Figure~\ref{1}. Others include the North Atlantic basin, the North-East Pacific basin, the North-West Pacific basin, the North Indian ocean basin, the South-East Indian ocean basin (Australian basin) and the South Pacific Australian basin.

The South-West Indian basin extends from longitude (30$^{\circ}$E to 90$^{\circ}$E) and latitude ranging from the equator to a little below the tropic of Capricorn. It is bounded to the north by India, to the East by Southern Africa and to the West by Australia.

\subsubsection{Climatology of study area}
The spatial distribution of tropical cyclones over the South-West Indian ocean is usually between latitude 35$^{\circ}$E to 95$^{\circ}$E, cutting across Madagascar and touching some parts of southern Africa.

\begin{figure}[!h]
% Use "\centering" in floats (figure, table), but if you need to center
% some text (why?) use "\begin{center}...\end{center}".
\centering 
% Figure environments same as 0.8 * \textwidth please
% That does not necessarily mean the actual picture size,
% it is a guideline for the environment which could contain
% 2 or more pictures! Be consistent and follow the guidelines
% provided in your sources.
\includegraphics[width=0.6\textwidth]{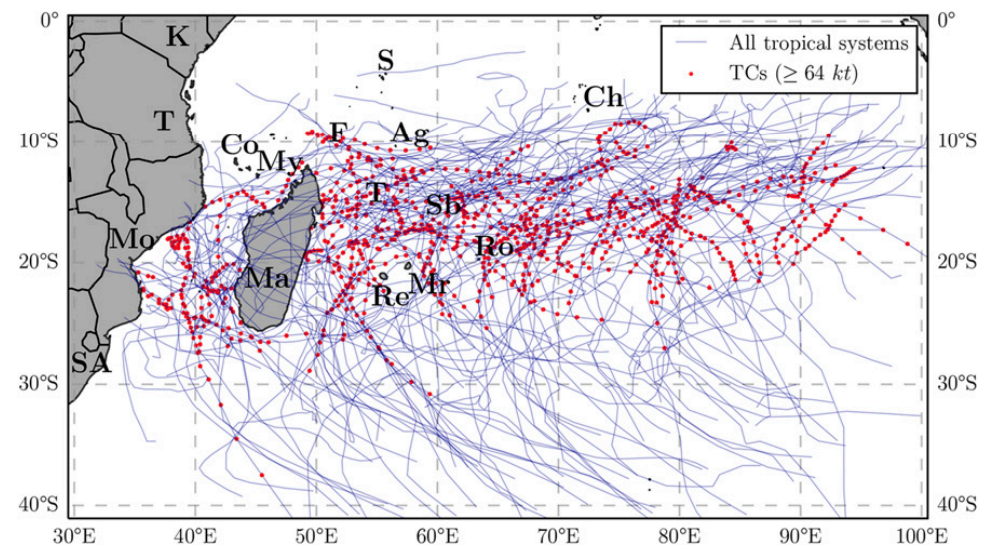}
\caption{ Map showing the tracks of tropical systems in the South-West Indian ocean with a VMAX observation during the 1999–2016 period. Tropical systems which reached a VMAX of greater than 64kt are spotted in red colours . Letters SA, Mo, T, and K indicate South Africa, Mozambique, Tanzania, and Kenya, respectively. Letters Co, My, Ma, F, T, S, Re, Mr, Ag, Sb, Ro,
and Ch indicate the islands of Comoros, Mayotte, Madagascar, Farquhar, Tromelin, Seychelles,
La Réunion, Mauritius, Agalega, Saint Brandon, Rodrigues, and the Chagos Archipelago,
respectively \cite{doi:10.1175/JAMC-D-17-0094.1}}
\label{2} 
% if you move the label it breaks the reference numbering; 
% always have it *after* the caption.
\end{figure}

Moving with respect to the easterly trading wind, tropical cyclones in this ocean basin follow a southwestward track before curving about their initial track in a southward direction where they transcend from tropical to extratropical cyclones \cite{holland1983tropical}. The curve in track by cyclones in this region can also follow a south eastward track as shown in Figure~\ref{2}

More intense cyclone activities occur between 10$^{\circ}$S and 20$^{\circ}$S for the South-West Indian ocean basin (Figure~\ref{2}). However between longitude 30$^{\circ}$E and 45$^{\circ}$E (with the channel separating Mozambique and Madagascar), there is a shift in the location of the more intense tropical cyclone activities to between 15$^{\circ}$S and 30$^{\circ}$S.
This displacement in the position of the more intense cyclones could be associated to higher sea surface temperatures within  the Mozambique channel, as it has known that tropical cyclones require warm SST (from Figure~\ref{2})  to originate and be sustained.

The intensity of tropical cyclones over the South-West Indian ocean is quite similar to those of hurricanes and typhoons found in the North-East and North-West Pacific ocean which have a cooler SST as shown in Figure~\ref{3}
%\newpage
\begin{figure}[!h]
% Use "\centering" in floats (figure, table), but if you need to center
% some text (why?) use "\begin{center}...\end{center}".
\centering 
% Figure environments same as 0.8 * \textwidth please
% That does not necessarily mean the actual picture size,
% it is a guideline for the environment which could contain
% 2 or more pictures! Be consistent and follow the guidelines
% provided in your sources.
\includegraphics[width=0.75\textwidth]{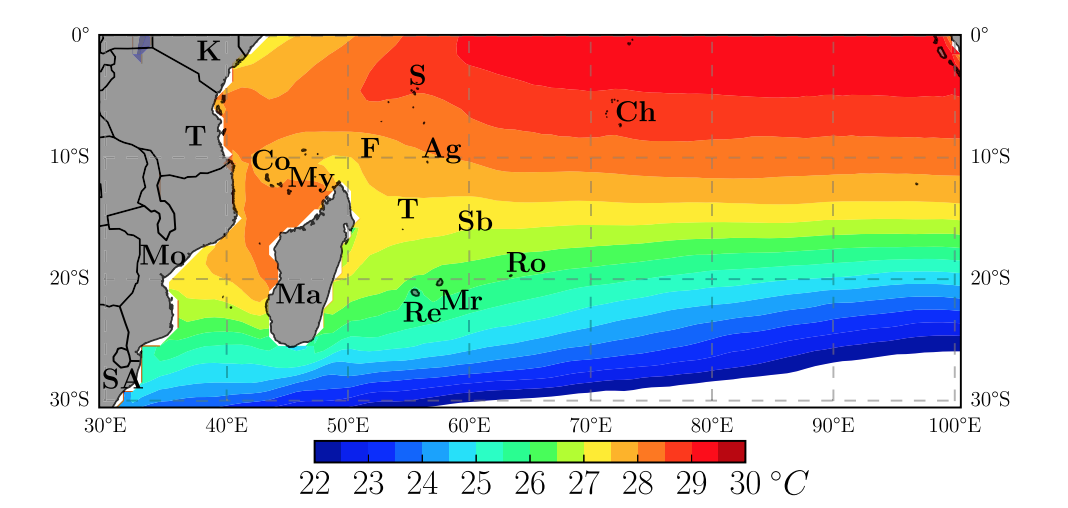}
\caption{. Climatology of mean SSTs exceeding 22$^{\circ}$C in the meridional part of the SWIO basin
for the active cyclone seasons from September to June during the 1999–2016 period. Temperatures are extracted from the 6-h-resolution global EI dataset at 0.75$^{\circ}$ latitude–longitude
resolution \cite{dee2011era}. Letters indicate the various islands or countries as in Figure~\ref{chi} The
total sample size is 170 months. }
\label{3} 
% if you move the label it breaks the reference numbering; 
% always have it *after* the caption.
\end{figure}

\section{Literature Review}

\subsection{Characteristics of cyclones over South West Indian ocean}
The nature of tropical cyclogenesis over the South-West Indian ocean has not received enough attention as that of the North Atlantic basin or any other basin \cite{leroux2018climatology}. Tropical cyclones in this region appears to be the least studied in peer-reviewed literature \cite{griffinpreliminary}. This is because, the relative absence of land around this region would imply that tropical cyclones systems which originate in this region would rarely make landfall, and in the rare ocassion that they do, the countries affected (Madagascar, Mozambique) do not have a vibrant meterological scientific community.
Most tropical systems which occur in this region affect vulnerable islands such as Madagascar (>22 million inhabitants) known for it's agro-based economy or the Mascarene Islands, which include La Réunion (870,000 inhabitants) and Mauritius (1.3 million inhabitants).

In the Southern Hemisphere, a single season of tropical cyclone goes from the 1st of July through to 30th of June. This corresponds with the season of the Northern Hemispheric equivalent \cite{griffinpreliminary}. However, historical data emphasizes that the activities of tropical cyclones in the South-West Indian
Ocean (0–40$^{\circ}$S, 30–90$^{\circ}$E), tropical storms (including tropical cyclones) occur mainly from late November to April, with on average 10 per season which usually reaches a peak during the months of January and February \cite{mavume2009climatology,malan2013variability}.

Over the past decade, much emphasis on modelling perspectives has been given to the process in which tropical cyclones intensify and the factors influencing this. However, there is still a lack of understanding of the intensification and hence forecasting of cyclone intensity has been relatively uncertain compared to forecast of cyclone tracks \cite{demaria1994statistical}.
Tropical cyclones Eline and Favio are examples of known cases which occured in the last decade around the South West Indian ocean  where there was a sudden intensification of the storms, causing severe damage in property and loss of lives in Madagascar and Mozambique \cite{reason2004tropical,klinman2008peculiar}. Intensifications of this nature usually occurs when the cyclone passes over regions of positive upper-ocean thermal anomalies \cite{scharroo2005satellite,lin2009warm}.

\subsubsection{Cyclone Enawo}

Cyclone Enawo was an intense tropical cyclone which struck Madagascar in March 2017. It was identified as the strongest  cyclone to ever hit Madagascar after cyclone Gafilo in 2004. Cyclone Enawo killed a total of 81 people amongst other property that were destroyed.

The cyclone started off as a moderate tropical depression on the 3rd of March after which it drifted slowly while intensifying. On the 5th of march, it intensified into a tropical cyclone and on the 6th of march, it transcended into an intense tropical cyclone. On the 7th of march, cyclone Enawo made landfall over the Sava region in Madagascar just after attaining it's peak intensity with a 10 minute sustained windspeed of 205$kmh^{-1}$ (Figure~\ref{3}). On the 9th of march, it re-emerged from the land back into the Indian ocean as a tropical depression moving South West - South, over central and Southern Madagascar, weakening into a
Tropical Depression, but still damaging infrastructures, causing floods and landslides in
several areas of the country \cite{probst2017tropical}.
The damage caused by cyclone Enawo was was felt by several municipalities of Madagascar, mainly Antalaha and Maroantsetra.

In the United nations (UN) Situation Report nr. 5 on 14 of April 2017, estimated, economic losses conducted by the CPGU (Cellule de Prevention et de Gestion des Urgences) and the World Bank was: \$400 million (about 4\% of annual GDP of Madagascar).

\begin{figure}[!h]
% Use "\centering" in floats (figure, table), but if you need to center
% some text (why?) use "\begin{center}...\end{center}".
\centering 
% Figure environments same as 0.8 * \textwidth please
% That does not necessarily mean the actual picture size,
% it is a guideline for the environment which could contain
% 2 or more pictures! Be consistent and follow the guidelines
% provided in your sources.
\includegraphics[width=0.5\textwidth]{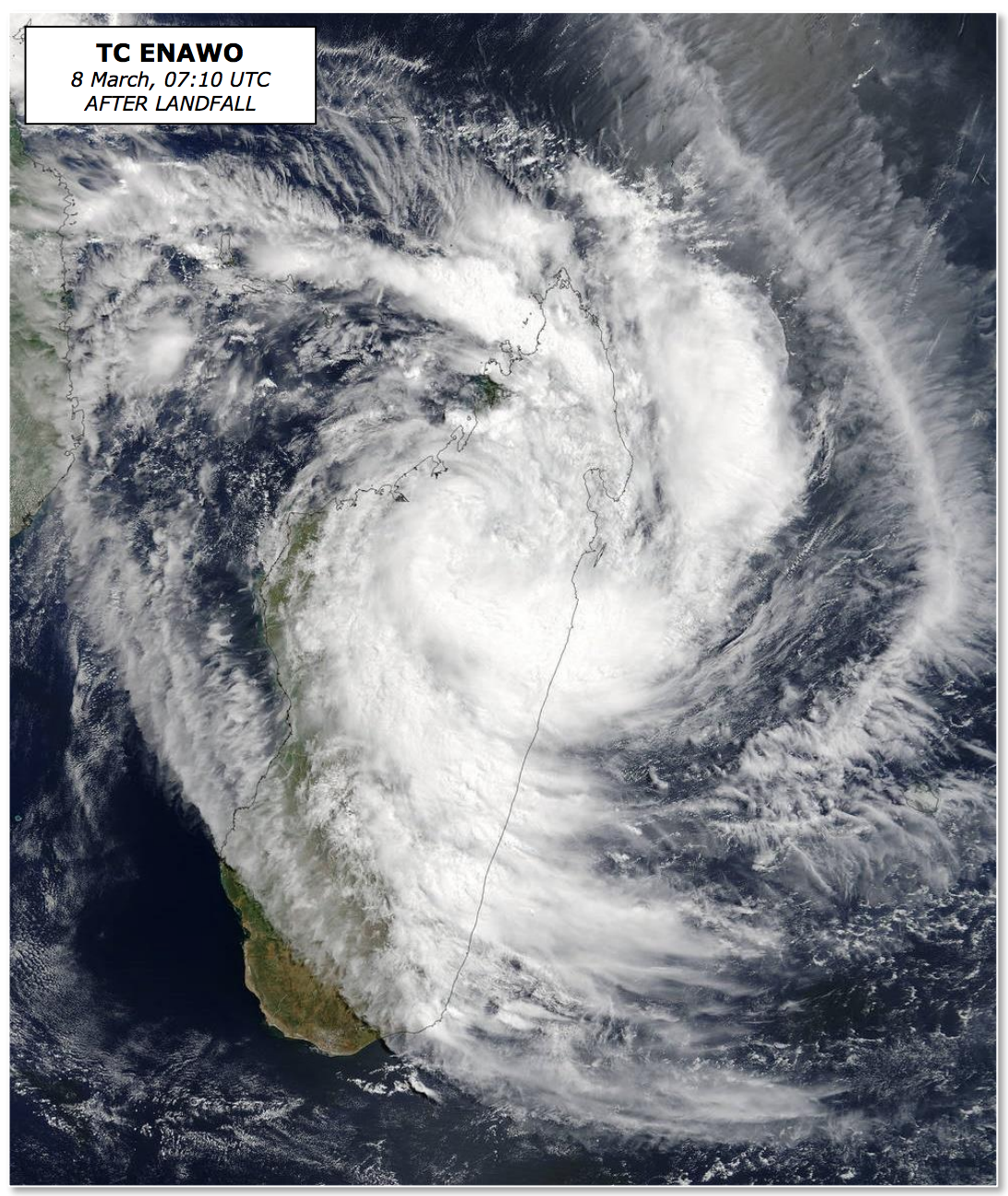}
\caption{Tropical cyclone Enawo at peak intensity  just before it makes landfall over Madagascar on the 7th of March 2017\cite{nass} }
\label{3} 
% if you move the label it breaks the reference numbering; 
% always have it *after* the caption.
\end{figure}

\begin{figure}[!h]
% Use "\centering" in floats (figure, table), but if you need to center
% some text (why?) use "\begin{center}...\end{center}".
\centering 
% Figure environments same as 0.8 * \textwidth please
% That does not necessarily mean the actual picture size,
% it is a guideline for the environment which could contain
% 2 or more pictures! Be consistent and follow the guidelines
% provided in your sources.
\includegraphics[width=0.6\textwidth]{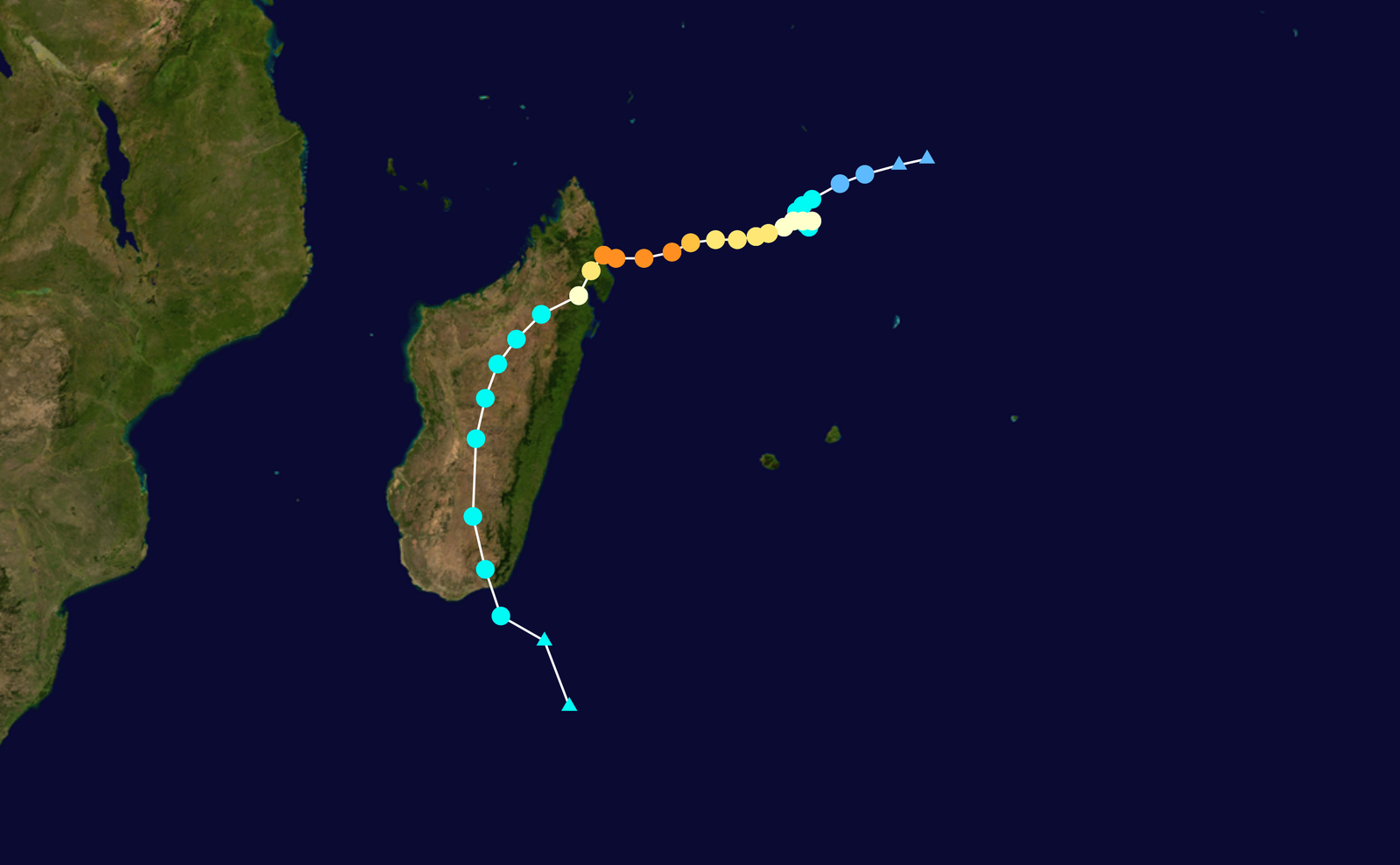}
\caption{The track followed by tropical cyclone Enawo from it's origin till it's dissipation \cite{fass}}
\label{xxxx} 
% if you move the label it breaks the reference numbering; 
% always have it *after* the caption.
\end{figure}

\subsection{Impact of climate change on tropical cyclones}

The importance of fluctuations in the intensity and frequency of tropical cyclone activities can not be over emphasized, especially since the affected population have been on a steady increase \cite{lamba}.
Amongst other natural disasters, tropical cyclones has been estimated to be one of the costliest in terms of loss of lives and property in the United states \cite{jamba}.
Furthermore, recent research proposes the importance of the activities of tropical cyclones in the regulation of thermohaline circulation in the oceans \cite{kamba}. This has an important feedback effect on regional and global climate.

One of the major challenges in attributing activities of tropical cyclones with climate change is to establish whether or not the changes in cyclone behaviour is not due to short term climate changes known as climate variability \cite{knutson2010tropical}. Only after this, can we attribute this to climate change.

According to previous theoretical research and practical observations done in the past, the intensity of tropical cyclones have been forecasted to increase with increasing global mean temperature \cite{emanuel1987dependence,knutson2004impact}.
However, these trend-detection in cyclone activities with temperature has been more focussed on the frequency.
The theoretical intensity that tropical cyclones could attain was modelled and hence proposed that the presence of a climate affected by greenhouse gases would impact tropical cyclones by increasing their potential intensities \cite{emanuel1987dependence}. Since then, alternative potential intensity theories has provided some support to this perspecive \cite{ja}.

\begin{figure}[!h]
% Use "\centering" in floats (figure, table), but if you need to center
% some text (why?) use "\begin{center}...\end{center}".
\centering 
% Figure environments same as 0.8 * \textwidth please
% That does not necessarily mean the actual picture size,
% it is a guideline for the environment which could contain
% 2 or more pictures! Be consistent and follow the guidelines
% provided in your sources.
\includegraphics[width=0.8\textwidth]{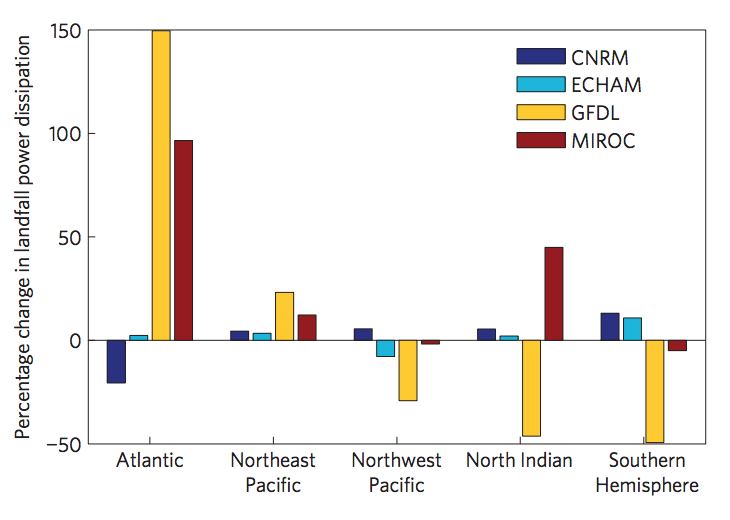}
\caption{ Effect of climate change on the power of tropical cyclones as shown by four climate models \cite{knutson2010tropical}}
\label{dent} 
% if you move the label it breaks the reference numbering; 
% always have it *after* the caption.
\end{figure}

The most commonly used metric for essessing the intensity of tropical cyclones is its \textit{power.} This is expressed as the cubed cumulative wind speed of each storm over track it traverses. Tracks followed by cyclones differ across models, hence the power estimated. Figure~\ref{dent} displays the effect of climate change on the power of tropical cyclones in different basins as described by four different climate models.
Notice that the power in most basins show an alternating increase and decrease except in the North-Pacific basin which shows a consistently increasing power.

\subsection{Modelling of tropical cyclones}

Tropical cyclones are described as mesoscale scale structures with a synoptic scale powerplant \cite{chia}. This implies that they are controlled by a variety of atmospheric processes. A fairly accurate model of tropical cyclones needs to account for cloud convection, sea surface temperature, wind speed amongst other processes. As such, general climate models are not suitable for modelling of tropical cyclones because fine processes are not well captured in it's coarse grids. This is in contrast with regional climate models whose higher resolution allows for these.

In modelling the potential risk of tropical cyclones, more attention is payed to it's tendency to make landfall. Therefore statistical models are built using exlusive historial data where tropical cyclones made landfall \cite{chill}.
However, comparatively there are few such cases where hurricanes made landfall, thus risk assessment is somewhat difficult. This limitation of data scarcity is overcome by using historical data for cyclone which never made landfall as well, thus increasing the amount of data to be used in the contruction of the statistical model.

One of the most important considerations in modelling a tropical cyclone across a basin is it's track from genesis to dissipation. Apart from the track, cyclone intensity are also studied on a basin-wide scale. Models like the autoregressive model uses increments for direction and speed forecast with random error terms \cite{chinko}.
Autoregressive model that works quite differently by using latitude and longitude increments as opposed to velocity can also be explored \cite{elle}.

Other modelling paradigm exists such as the three-dimensional hurricane modeling studies using regional nested modeling approaches. However, an identified limitation of this model is that it makes use of the current climate conditions as input, obtained from single global climate model—the Geophysical Fluid Dynamics Laboratory (GFDL) R30 coupled model \cite{knutson2004impact}.

\section{Dataset and Methodology}

\subsection{Model description}
The model used for this study is the Unified
Environmental Modeling System (UEMS) - Weather Research Forecast (WRF) model regional climate model.
It is a complete, full-physics, state-of-the-science numerical weather prediction (NWP) package that incorporates the the National Oceanic and Atmospheric Administration’s (NOAA) and WRF systems into a single user-friendly, end-to-end forecasting system. 

\begin{figure}[!h]
% Use "\centering" in floats (figure, table), but if you need to center
% some text (why?) use "\begin{center}...\end{center}".
\centering 
% Figure environments same as 0.8 * \textwidth please
% That does not necessarily mean the actual picture size,
% it is a guideline for the environment which could contain
% 2 or more pictures! Be consistent and follow the guidelines
% provided in your sources.
\includegraphics[width=0.7\textwidth]{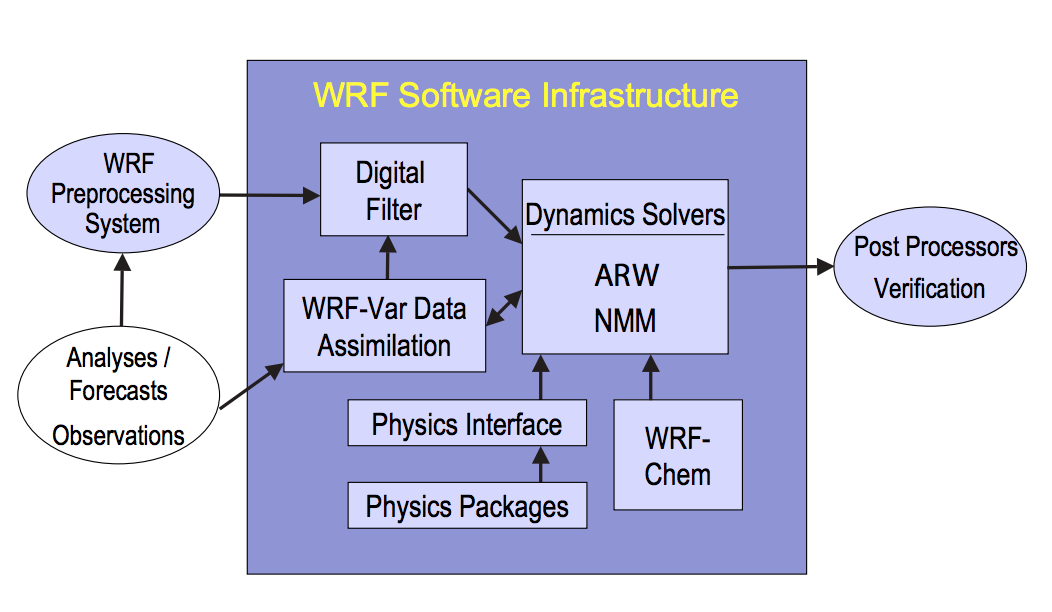}
\caption{Components of the WRF system. \cite{mlle} }
\label{li} 
% if you move the label it breaks the reference numbering; 
% always have it *after* the caption.
\end{figure}

Figure~\ref{li} shows the components of the WRF regional climate model and it's data processing/forecasting workflow. The WRF framework includes components for dynamical solvers, initialization, WRF-Var, WRF-Chem amongst others.
The two options for dynamical solvers in the WRF model includes: the Advanced Research WRF (ARW) solver developed primarily at  National Center for Atmospheric Research (NCAR), and the Nonhydrostatic Mesoscale Model (NMM) solver developed at National Centers for Environmental Prediction (NCEP).

\quad The Weather Research and Forecasting (WRF) model is a numerical weather prediction (NWP)
and atmospheric simulation system designed for both research and operational applications \cite{mlle}.
It's developement was a collaborative effort by NCAR, Atmospheric Administration’s
(NOAA), the Department of Defense’s Air Force Weather Agency (AFWA) and the Center for Analysis and Prediction of Storms (CAPS) to develop a mesoscale forecasting system capable of predicting mesoscale weather and drive research opportunities into it's operations.

The ARW is one of two dynamical solvers which are a subset of the WRF model. The
ARW solver combines with the NMM solver shown in Figure~\ref{li} and all other WRF components within the framework. It includes real-data and idealized simulations, various lateral boundary condition options, full physics options, non-hydrostatic and hydrostatic (runtime option) amongst others. 
There are no significant differences between the designs of the NMM and ARW dynamical cores except for the output of model physics \cite{mlle}. The NMM core is a fully compressible, hydrostatic NWP which was later extended to include non-hydrostatic options. On the other hand, the ARW is a fully compressible, Eulerian, non-hydrostatic model with run-time hydrostatic option.

For the purpose of interpretability in the equations below,ignoring the Coriolis effect alongside variables which account for the earths curvature and diffussion,the WRF model can be configured to solve the following equations:

%Equations with numbering
\begin{align}
\frac{d \mathbf{v}}{d t}=-\frac{1}{\rho} \nabla p-g \mathbf{k}
\end{align} 

%Equations with no numbering in specific line by using \nonumber

\begin{align}
\frac{d \rho}{d t}=-\rho \nabla \cdot \mathbf{v}
\end{align} 

\begin{align}
\frac{d \theta}{d t}=\frac{Q}{C_{p} \pi}
\end{align} 

\begin{align}
p=\rho R T
\end{align} 

The above equations represents the momentum equation (Navier-Stokes), continuity equation, thermodynamic equation and the equation of state respectively.
Where \textbf{v} is the wind vector  representing $(u, v, w)$ as the velocity components in the $(x, y, z)$ directions, \textbf{g} is the acceleration due to gravity, $R$ represents the molar gas constant for dry air, $Q$ is the heating due to adiabatic process, $C_p= \left(\frac{7}{2}\right) \times R $ represents the specific heat capacity of dry air, while \textbf{k} represents the unit vector in the vertical direction.
$\pi$ and $\theta$ denotes the exner function and the potential temperature respectively and are defined as:

\begin{align}
\pi=\left(\frac{p}{p_{0}}\right)^{R / C p} \text{ and } \theta=\frac{T}{\pi}
\end{align} 

When solved, these equations estimate the state of dry air/atmosphere expressed in terms of temperature $(T)$, pressure $(p)$, density $(\rho)$ and the windspeed components $(u, v, w)$.
However, the run-time hydrostatic option for the ARW assumes hydrostatic balance for the momentum equation in the $w$ direaction and thus represents motion in the vertical direction as a balance between the pressure gradient force (PGF) and the gravitational pull, described mathematically as:

\begin{align}
\frac{\partial p}{\partial z}=-\rho g
\end{align}

The formulation of the model equations adapts a mass-based terrain coordinate system solved in Arakawa-C grid using Runge–Kutta third-order time integration techniques.
ARW modeling system supports horizontal nesting that allows resolution to be focused over a region of interest by introducing an additional grid (or grids) into the simulation with the choice of one-way and two-way nesting procedures. \cite{dodla2016prediction}

\subsection{Model configuration  \& data description}
\begin{figure}[!h]
% Use "\centering" in floats (figure, table), but if you need to center
% some text (why?) use "\begin{center}...\end{center}".
\centering 
% Figure environments same as 0.8 * \textwidth please
% That does not necessarily mean the actual picture size,
% it is a guideline for the environment which could contain
% 2 or more pictures! Be consistent and follow the guidelines
% provided in your sources.
\includegraphics[width=0.5\textwidth]{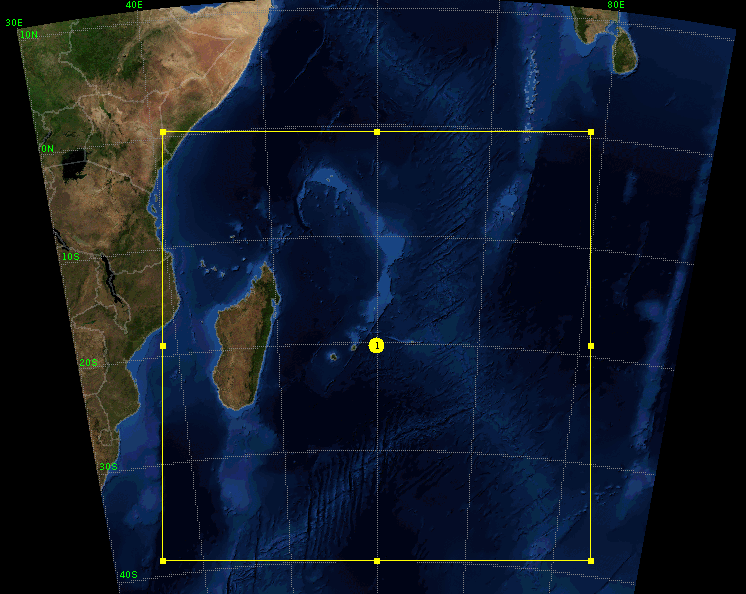}
\caption{Primary domain for WRF simulation}
\label{james} 
% if you move the label it breaks the reference numbering; 
% always have it *after* the caption.
\end{figure}

The ARW domain for the model simulation was defined as shown in Figure~\ref{james} below using the Lambert Conformal
Grid projection type for mid-latitudes centered over 0$^{\circ}$ longitude (GMT) with a centerpoint latitude of -20.211$^{\circ}$ and a centerpoint longitude of 59.798$^{\circ}$. The number of horizontal (NX) and vertical (NY) grid points chosen was 100 with an initial grid spacing of 45\textit{km} and a terrestial data resolution of 10 minute.

The domain bounded by the yellow box in Figure~\ref{james} spans latitude 40$^{\circ}$E to 80$^{\circ}$E and longitude 0$^{\circ}$N to 40$^{\circ}$S. 

After defining the domain, localization process of set domain was initialized. This serves to extract the terrestrial data over the area covered by the domain(s) from the large global files at the resolution(s) specified during your domain creation.

Following the localization of the domain, boundary conditions were downloaded from the National Center for Environmental Prediction (NCEP's) Climate Forecast System Reanalysis for my domain  at a 6 hour frequency for 10 days (March 2nd 2017 - March 11th 2017). The boundary conditions serve as input to the ARW solver in model simulation. The output of simulations were exported as NetCDF files. These consited of a total of 37 simulation results (4 simulations per day, for 10 days).

The choice of the domain for localization and dates in which boundary conditions were obtained was guided by the choice of the tropical cyclone to be studied.

\subsection{Experiment design}
\begin{figure}[!h]
% Use "\centering" in floats (figure, table), but if you need to center
% some text (why?) use "\begin{center}...\end{center}".
\centering 
% Figure environments same as 0.8 * \textwidth please
% That does not necessarily mean the actual picture size,
% it is a guideline for the environment which could contain
% 2 or more pictures! Be consistent and follow the guidelines
% provided in your sources.
\includegraphics[width=0.8\textwidth]{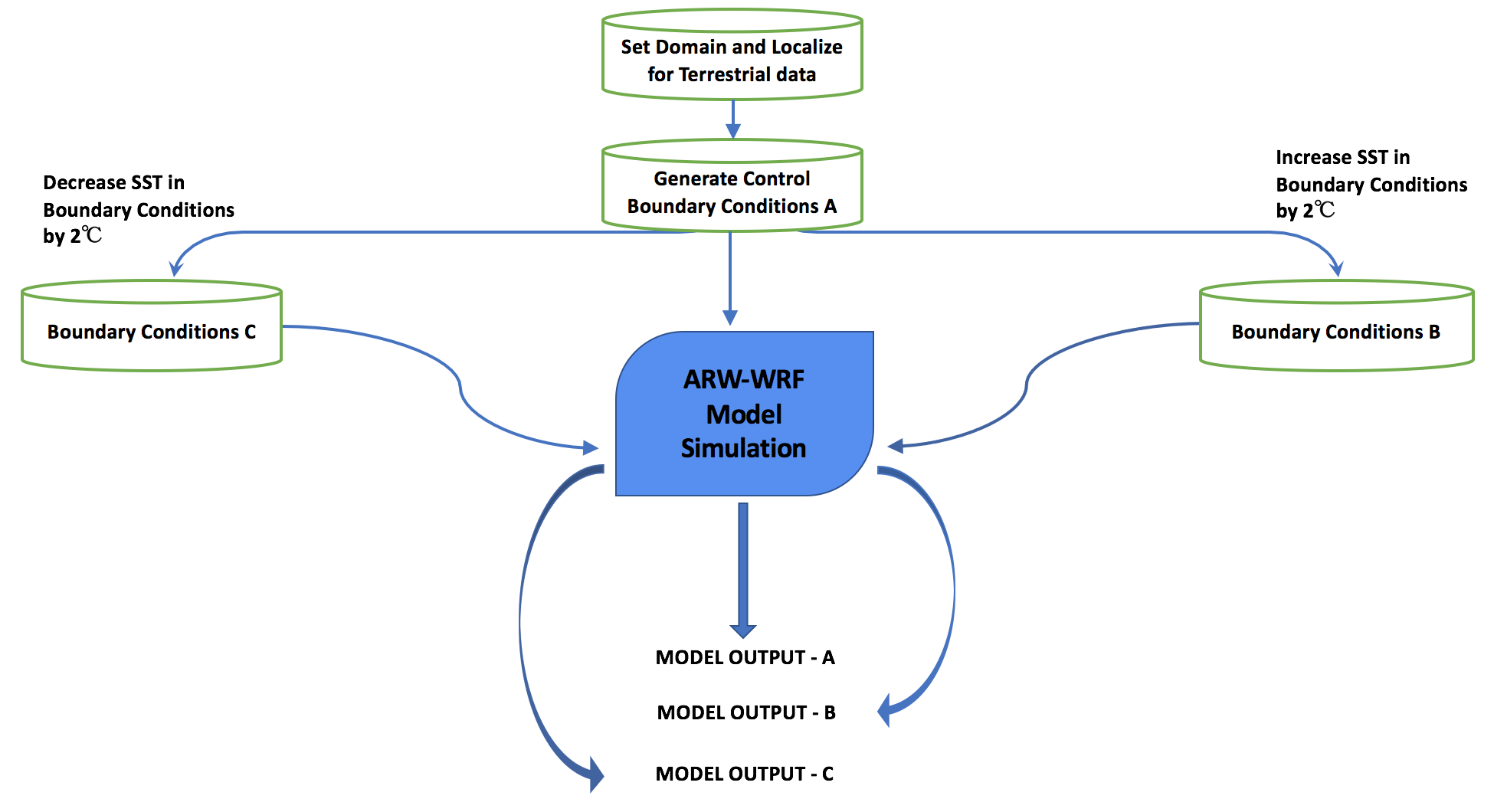}
\caption{Experiment design and workflow for observing the effect of sea surface temperature (SST) changes in boundary data on the track and intensity of tropical cyclone Enawo.}
\label{jame} 
% if you move the label it breaks the reference numbering; 
% always have it *after* the caption.
\end{figure}

Figure~\ref{jame} shows the experiment design workflow for investigating the influence of increasing/decreasing  sea surface temperature (SST) on tropical cyclone Enawo over the set domain. The experiment hinges on the increase and decrease of SST in the boundary condition dataset which serve as the input data for the ARW dynamical core of the WRF model.

After defining the domain for simulation, as well as the duration (2nd March 2017 to 11th March 2017) and frequency (4 hours per day) in which the boundary conditions are to be obtained, the following procedures were followed in this order for the purpose of observing the influence of temperature increase on the intensity of tropical cyclones within the domain.

\begin{itemize}

\item The boundary condition was used as input into the UEMS-WRF model in three phases:
	\begin{itemize}
		\item Unaltered boundary condition (Control) was inputed into the WRF model for simulation and the simulation was saved as the control output of the model. This boundary condition was labelled (A) as shown in Figure~\ref{jame}
		\item The sea surface temperature for the unaltered boundary condition above was increased by 2$^{\circ}C$, then inputed into the UEMS-WRF model. This boundary condition was labelled (B) as shown in Figure~\ref{jame}.
		\item The sea surface temperature for the unaltered boundary condition was then decreased by 2$^{\circ}C$ before inputed into the UEMS-WRF model for simulation. This boundary condition was labelled (C) as shown in Figure~\ref{jame}.

	\end{itemize} 
	
\item The output of the model from the three boundary conditions (A, B and C) were analysed comparatively to see if there was an increase/decrease in the intensity of the tropical cyclone Enawo with increased sea surface temperature.
Comparison of the intensity of the tropical cyclone was based on the following variables: the maximum precipitation rate, maximum windspeed, minimum surface pressure etc.

These variables were chosen amongst existing output variables from the WRF model simulation because they are variables that best identify intensity of tropical cyclones. Theoretically, the pressure  and windspeed in the eye of a tropical cyclone should be lower than the surrounding. Hence, we could perform a comparative analysis of the minimum pressure for all three of our simulation scenerios.

\end{itemize}

\section{Results and Discussion}

This section discusses the result of all of experiments ran using the UEMS-WRF model for the simulation of cyclone Enawo. The first section describes the validation of WRF model results with observed data from European Centre for Medium-Range Weather Forecasts (ECMWF) ERA5 re-analysis dataset, while the second section illustrates the sensitivity of various climatological parameters to increasing/decreasing SST.

\subsection{Model validation}
\subsubsection{Validating the Track of the Tropical Cyclone}
\begin{figure}[ht]
        \begin{subfigure}{.5\textwidth}
            \centering
            \includegraphics[scale=0.205]{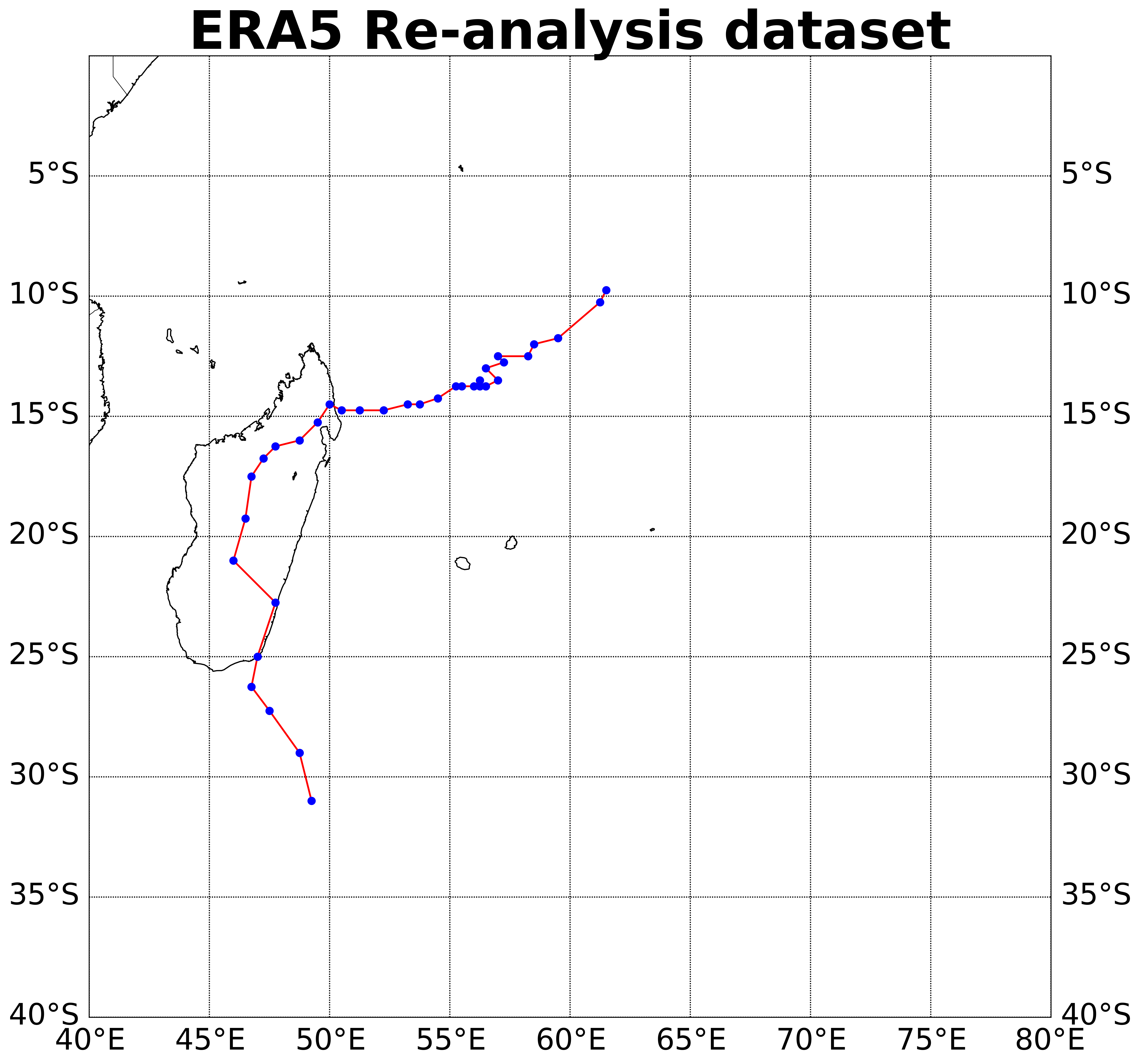}
        \end{subfigure}%
    \begin{subfigure}{.5\textwidth}
        \centering
\includegraphics[scale=0.21]{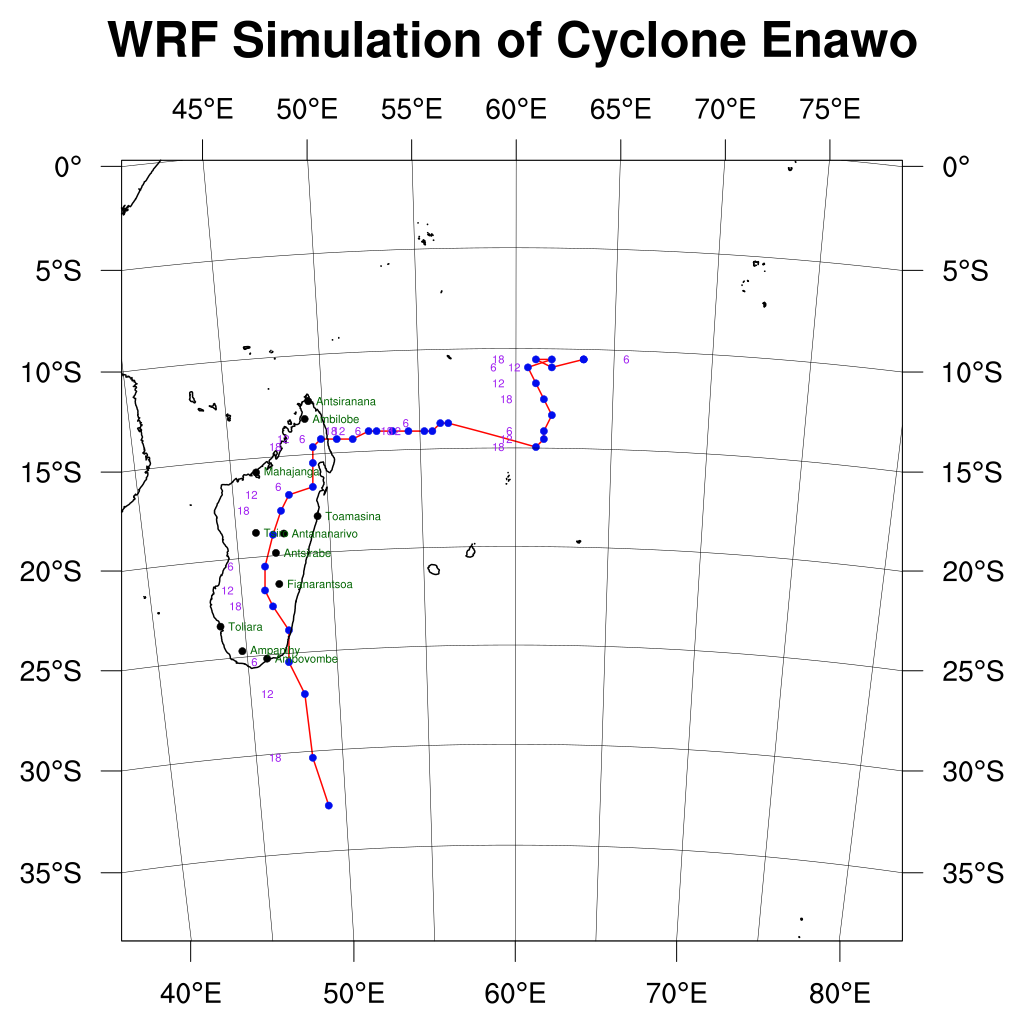}    
    \end{subfigure}%

\caption{A comparison of the observed track (a) for cyclone Enawo with the simulated track (b) by the WRF model. The WRF simulation of the track for cyclone Enawo uses it's low pressure vortex as parameter for 10 days (2nd March - 11th March, 2017). The NCL tool was used to locate the low pressure vortex used in identifying the track mapped from genesis of lysis.}
\label{liM} 
% if you move the label it breaks the reference numbering; 
% always have it *after* the caption.
\end{figure}

Figure~\ref{liM} (a) and (b) shows a comparison between the observed track (by meteo-france) for cyclone enawo with the track simulated by the WRF model. This shows that the WRF model reasonably reproduces the track for cyclone Enawo

Figure~\ref{liM} (b) shows the path followed by the tropical cyclone Enawo. This was simulated by monitoring the coordinates of the lowest sea-level pressure point as it drifts over time.
The simulation shows the low pressure core of tropical cyclone following an initial southward direction, before it changes to a western direction on the 4th of March 2017, where it pressure was about
 970 HPa. The cyclone continues on this western direction until the 7th of March, where the lowest pressure was recorded (957.624 HPa) just before it makes landfall.

After making landfall, the pressure gradually increases to about 988 HPa as it’s intensity decreases and it moves through cities like Antisabe, Fiamarantosa and Toasmania. The low pressure core of the cyclone moves in a southward direction through the land until it emerges at latitude 25$^{\circ}$S where it's intensity decreases as it’s pressure increases till it dissipates.

\newpage
\subsubsection{Validating the Windspeed Vector}
\begin{figure}[ht]
        \begin{subfigure}{.5\textwidth}
            \centering
            \includegraphics[scale=0.35]{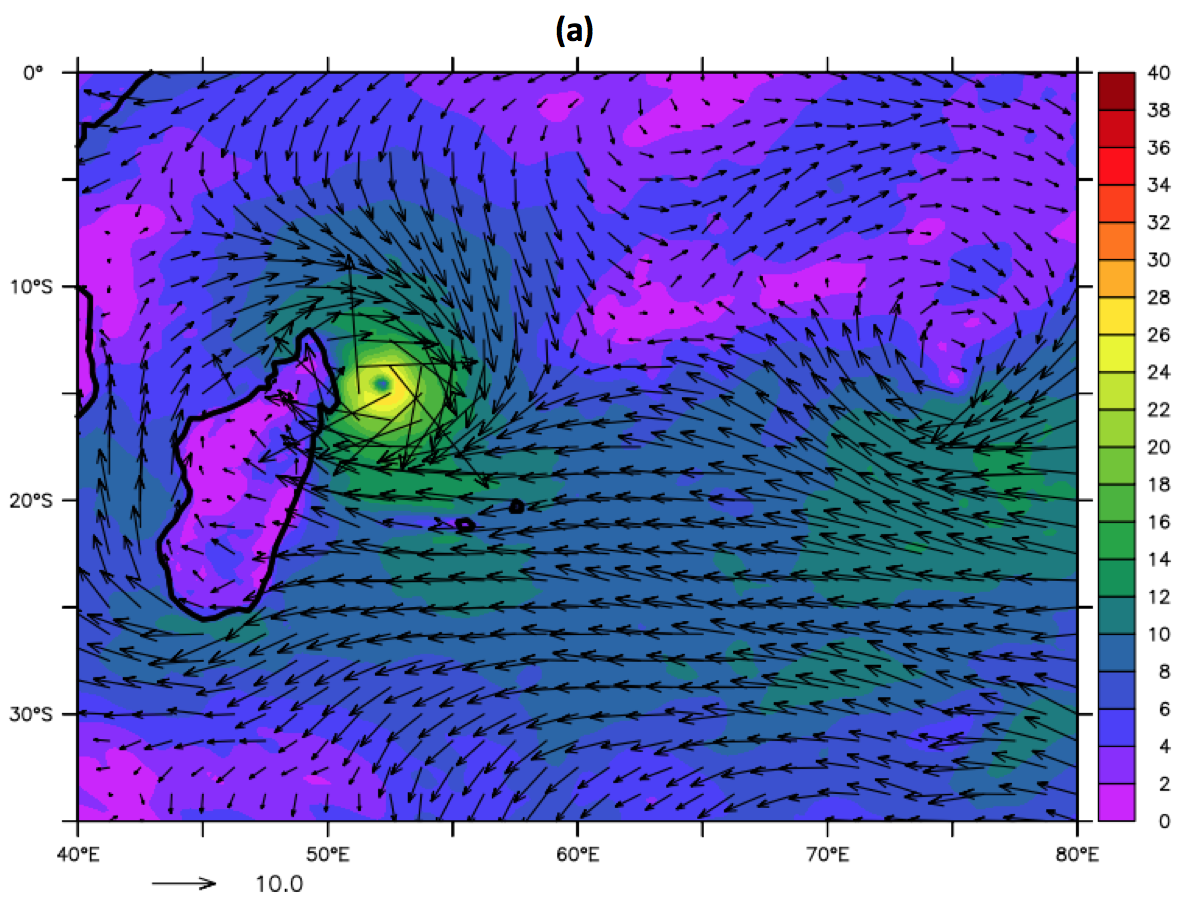}
        \end{subfigure}%
    \begin{subfigure}{.5\textwidth}
        \centering
\includegraphics[scale=0.35]{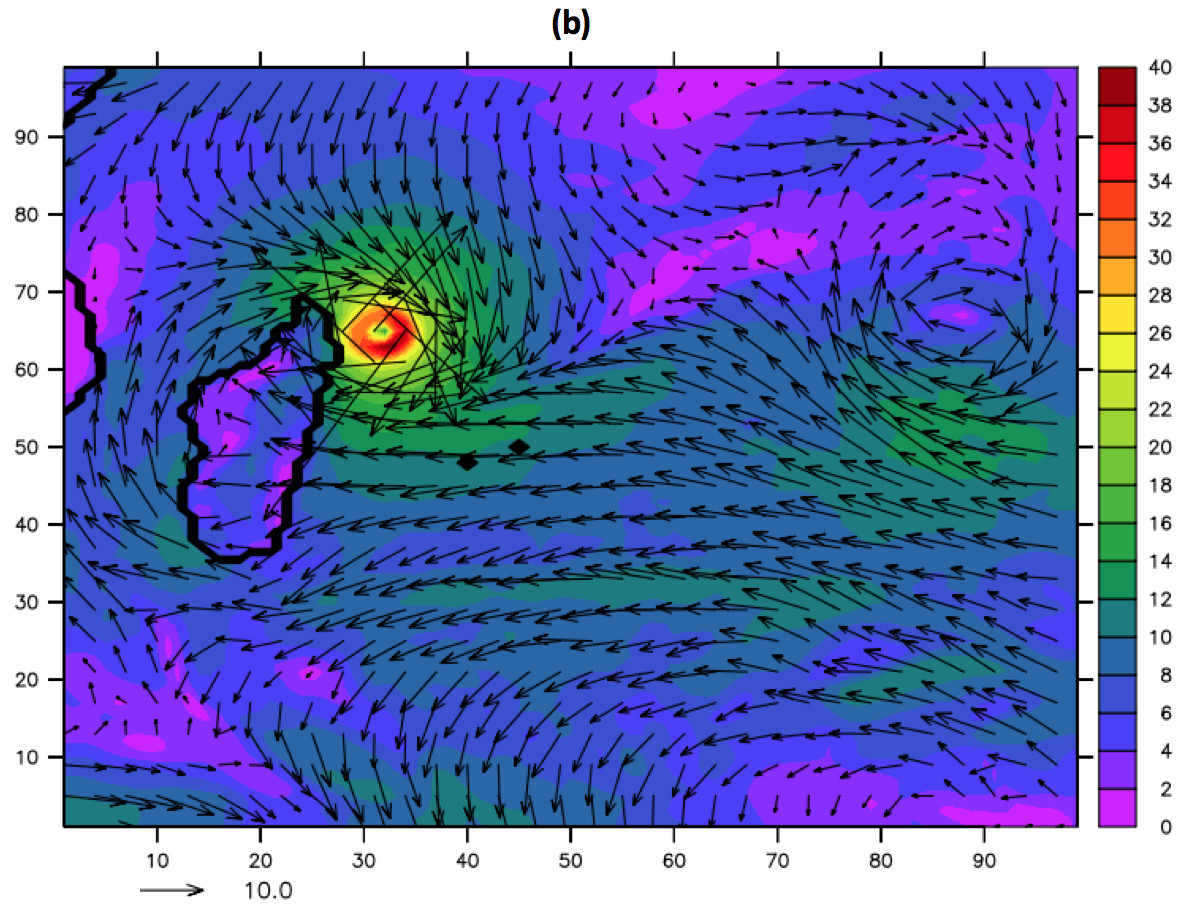}    
    \end{subfigure}%

\caption{Validation of the windspeed vector generated by the WRF model (b) with the European Centre for Medium-Range Weather Forecasts (ECMWF) ERA5 Reanalyses dataset (a) by the Copernicus climate change services, showing similarity in curl of vector fields.}
\label{jk} 
% if you move the label it breaks the reference numbering; 
% always have it *after* the caption.
\end{figure}

Figure~\ref{jk} (a) and (b) shows a validation of the output of windspeed vector from the WRF model simulation with the European Centre for Medium-Range Weather Forecasts (ECMWF) ERA5 reanalyses dataset by the Copernicus climate change services. This shows the similarity between the model output and observed data.

By comparison, we notice that two curls for velocity vector fields are located at longitude 52$^{\circ}$E and 75$^{\circ}$E on the European Centre for Medium-Range Weather Forecasts (ECMWF) ERA5 re-analysis data. However, these two vector curl fields are located at longitude 30$^{\circ}$E and 90$^{\circ}$E. 

The magnitude of the velocity field simulated by the WRF model is seen to be larger than for the observed European Centre for Medium-Range Weather Forecasts (ECMWF) ERA5 re-analysis data reanalysis data. Maxiumum windpspeed of about 40$ms^{-1}$ is identified in the WRF simulation of the cyclone location before it makes landfall. However, a maximum windspeed of about 26$ms^{-1}$ is noticed from the European Centre for Medium-Range Weather Forecasts (ECMWF) ERA5 re-analysis data. This suggests the WRF model exaggerated the windspeed during simulation. Also, in both the re-analysis data and simulated dataset, there is a noticeable low velocity core in the location for the tropical cyclone. This is consistent with the structure expected for tropical cyclones.

\newpage
\subsection{Sensitivity analysis}

\subsubsection{Sensitivity of Cyclone Track}
\begin{figure}[ht]
        \begin{subfigure}{.5\textwidth}
            \centering
            \includegraphics[scale=0.30]{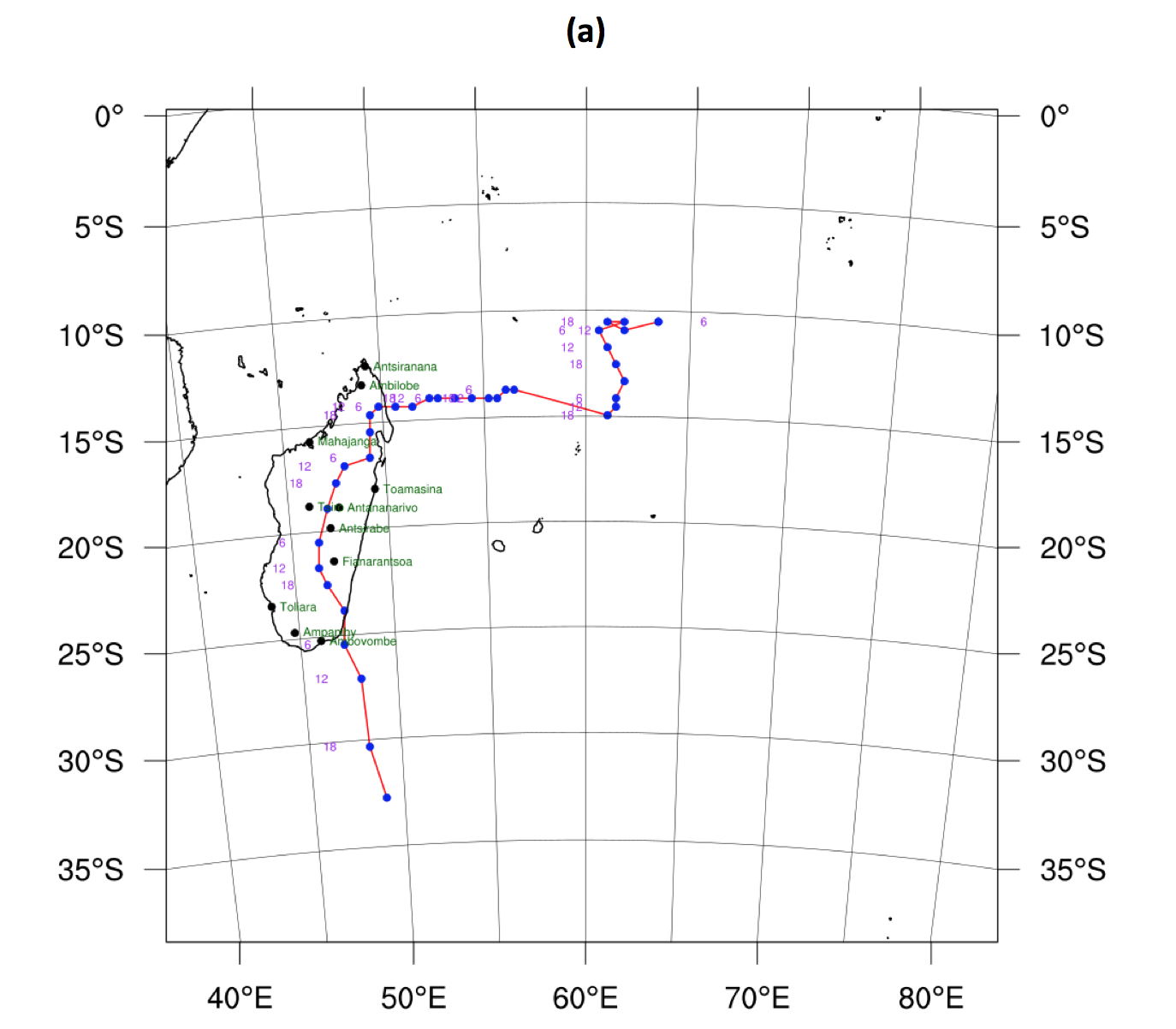}
        \end{subfigure}%
    \begin{subfigure}{.5\textwidth}
        \centering
\includegraphics[scale=0.30]{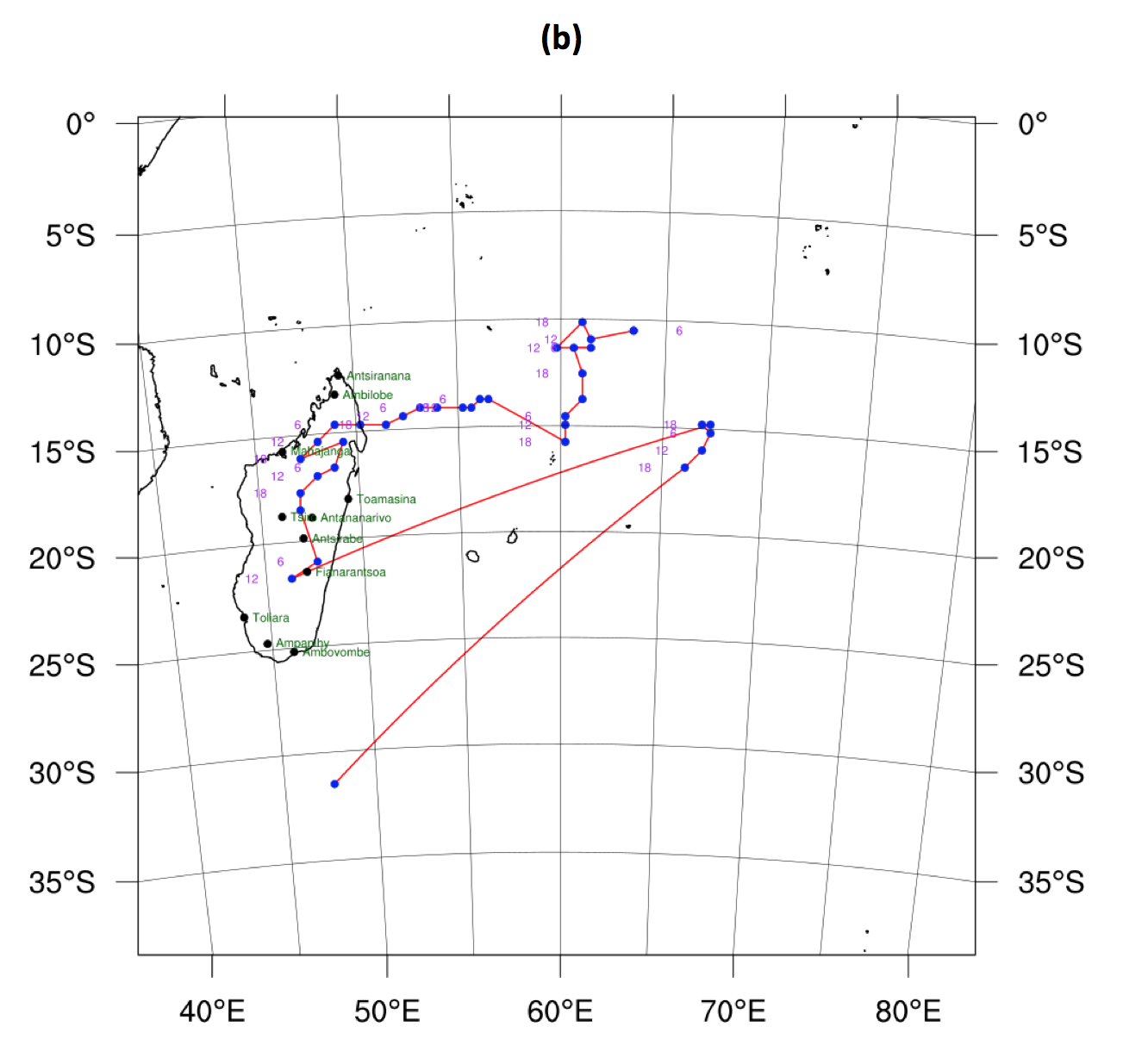}    
    \end{subfigure}%

    \begin{subfigure}{.5\textwidth}
        \centering
\includegraphics[scale=0.30]{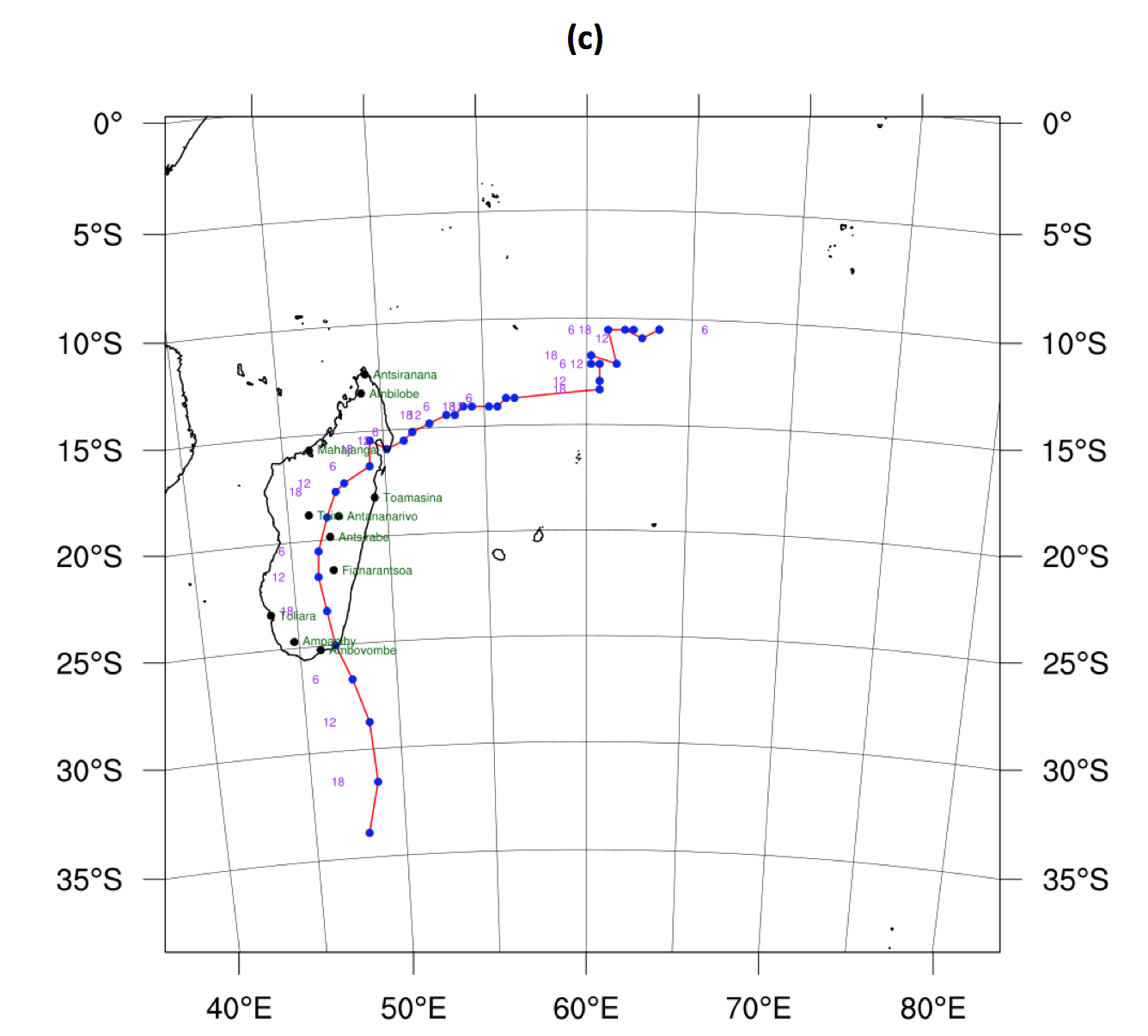}    
    \end{subfigure}%

    \caption{ Figure showing the track followed by cyclone Enawo from the 2nd of March to the 11th of march 2017. (a.) Control simulation, (b.) Simulation with SST increased by 2$^{\circ}C$. (c.) Simulation with SST decreased by 2$^{\circ}C$.}
    \label{comss}
\end{figure}

Figure~\ref{comss} (a), (b) and (c) below shows the effect of SST on the track followed by cyclone Enawo from it's genesis (2nd of March) to it's lysis (11th of March). We notice by comparing Figure~\ref{comss} (a) and (c), that decreasing the SST does not have a significant effect on the track followed by the cyclone, except for a more smoother trajectory.

On the other hand,comparing figure~\ref{comss} (a) and (b), there is a more noticeable distortion in the track of the cyclone with increased SST. Between the 2nd and 4th of March, there are minute haphazard tracks in the cyclone. The most noticeable phenomena is between the 10th to 11th of March, where a lower pressure system was identified towards the East. This is due to the fact that an increase in SST creates a another cyclone which occurs simultaneously as the previous one.

This suggests that an increase in SST has a potential of  increasing the number of cyclone occurences within a given time frame.

\newpage

\subsubsection{Sensitivity of Total Precipitation}
Figure~\ref{coms} below shows how the increase in SST temperature causes an increase in the total accumulated precipitation and it's spatial distribution, as compared to when the SST was unaltered and when it was decreased by 2$^{\circ}C$. 
Figure~\ref{coms} (a) shows the sum of the total precipitation for the beginning of the simulation period (2nd of March,2017) to the end of the simulation period (11th of March,2017).

  \begin{figure}[ht]
  \hspace{1.63in}
          \begin{subfigure}{.5\textwidth}
        \centering
\includegraphics[scale=0.28]{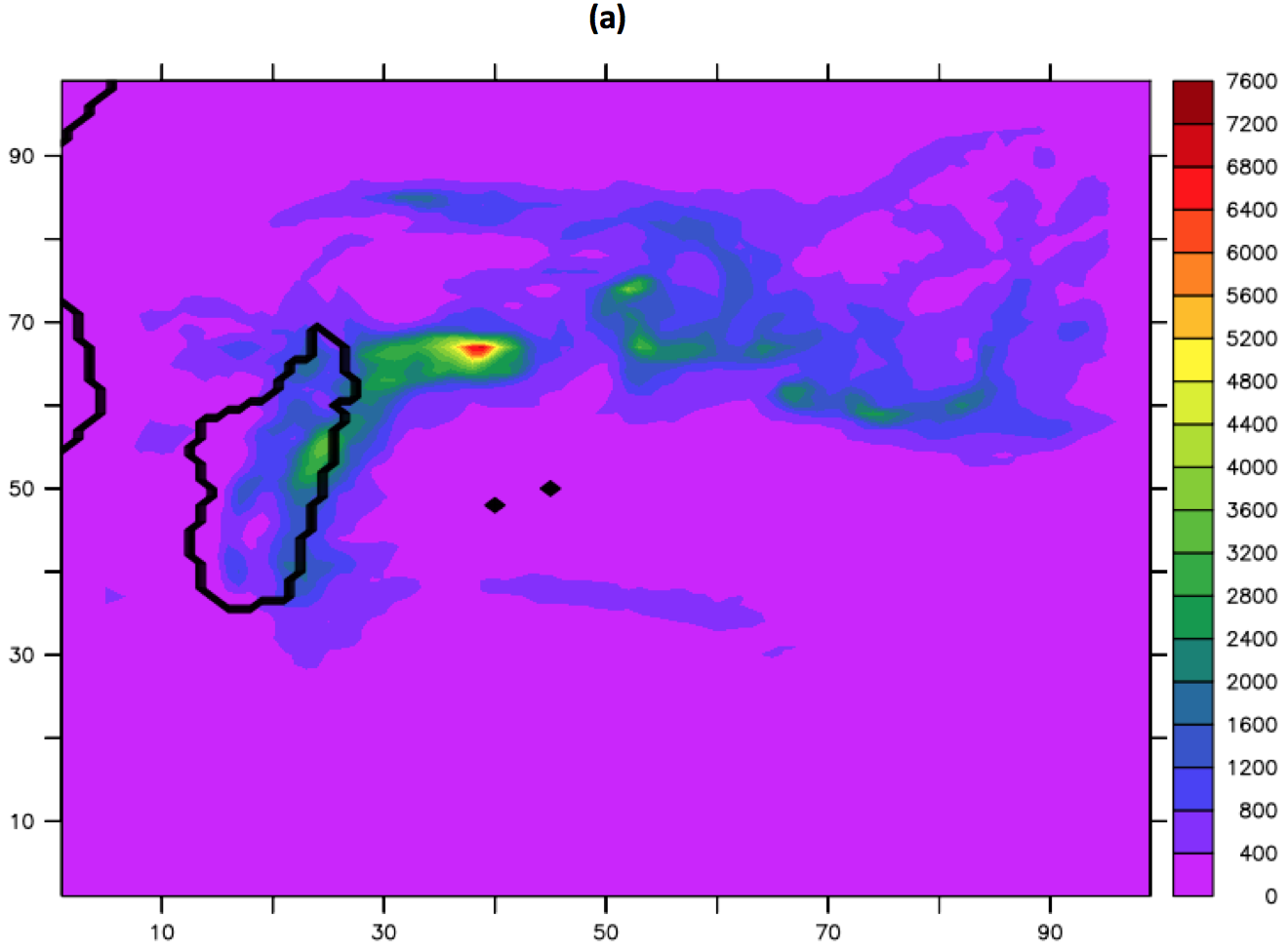}    
    \end{subfigure}%
    
        \begin{subfigure}{.5\textwidth}
            \centering
            \includegraphics[scale=0.28]{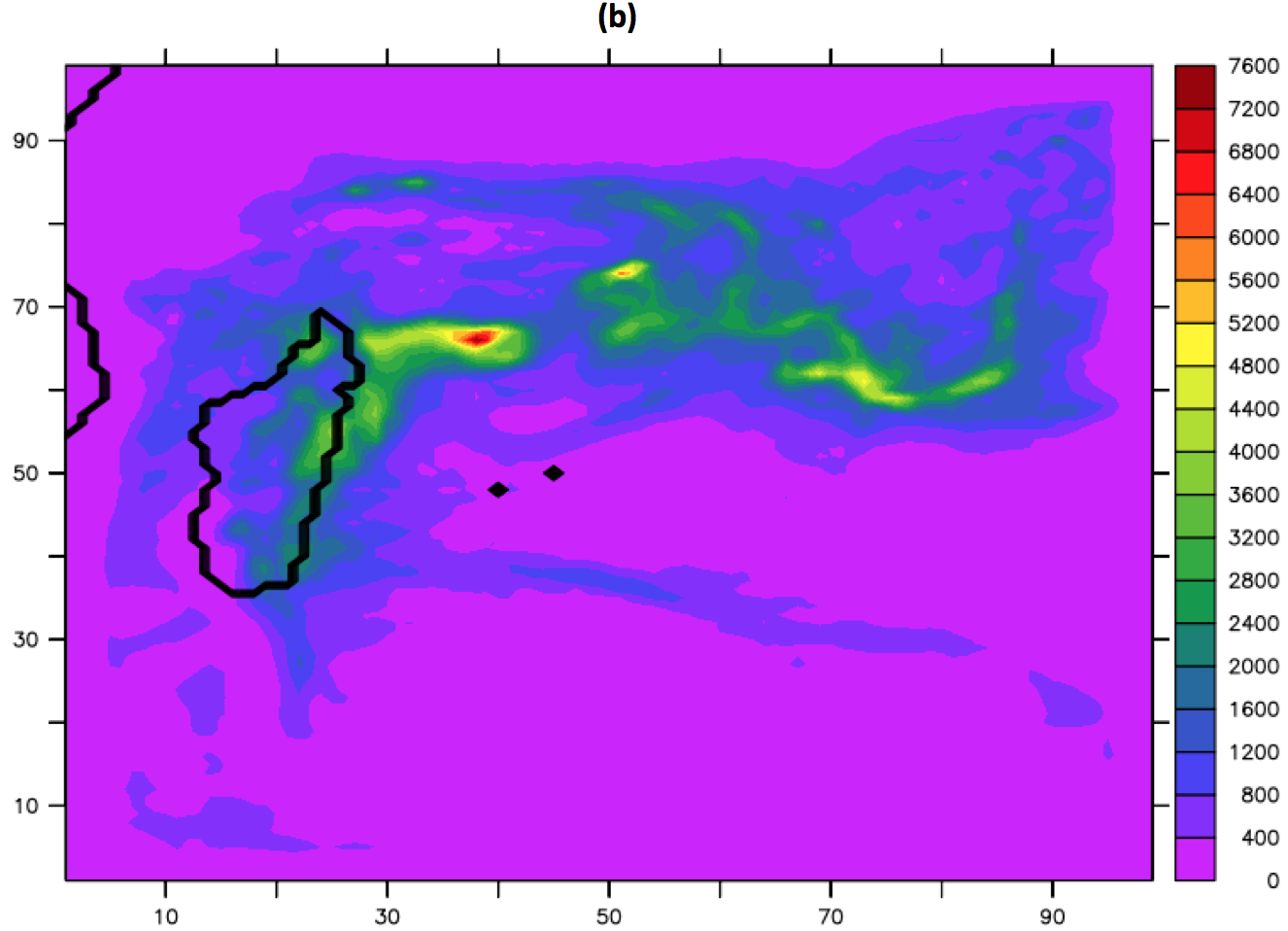}
        \end{subfigure}%
    \begin{subfigure}{.5\textwidth}
        \centering
\includegraphics[scale=0.28]{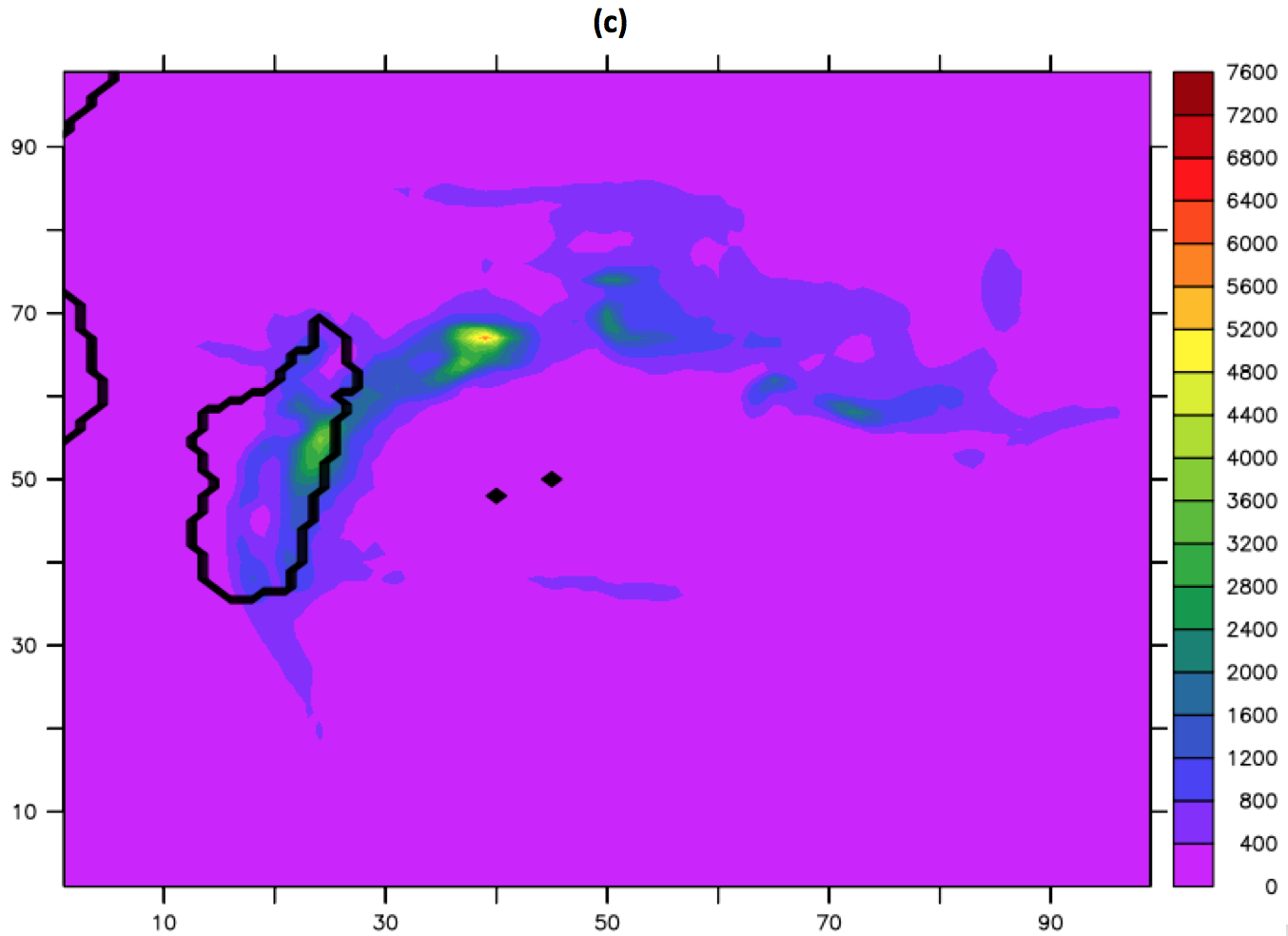}    
    \end{subfigure}%

    \begin{subfigure}{.5\textwidth}
        \centering
\includegraphics[scale=0.22]{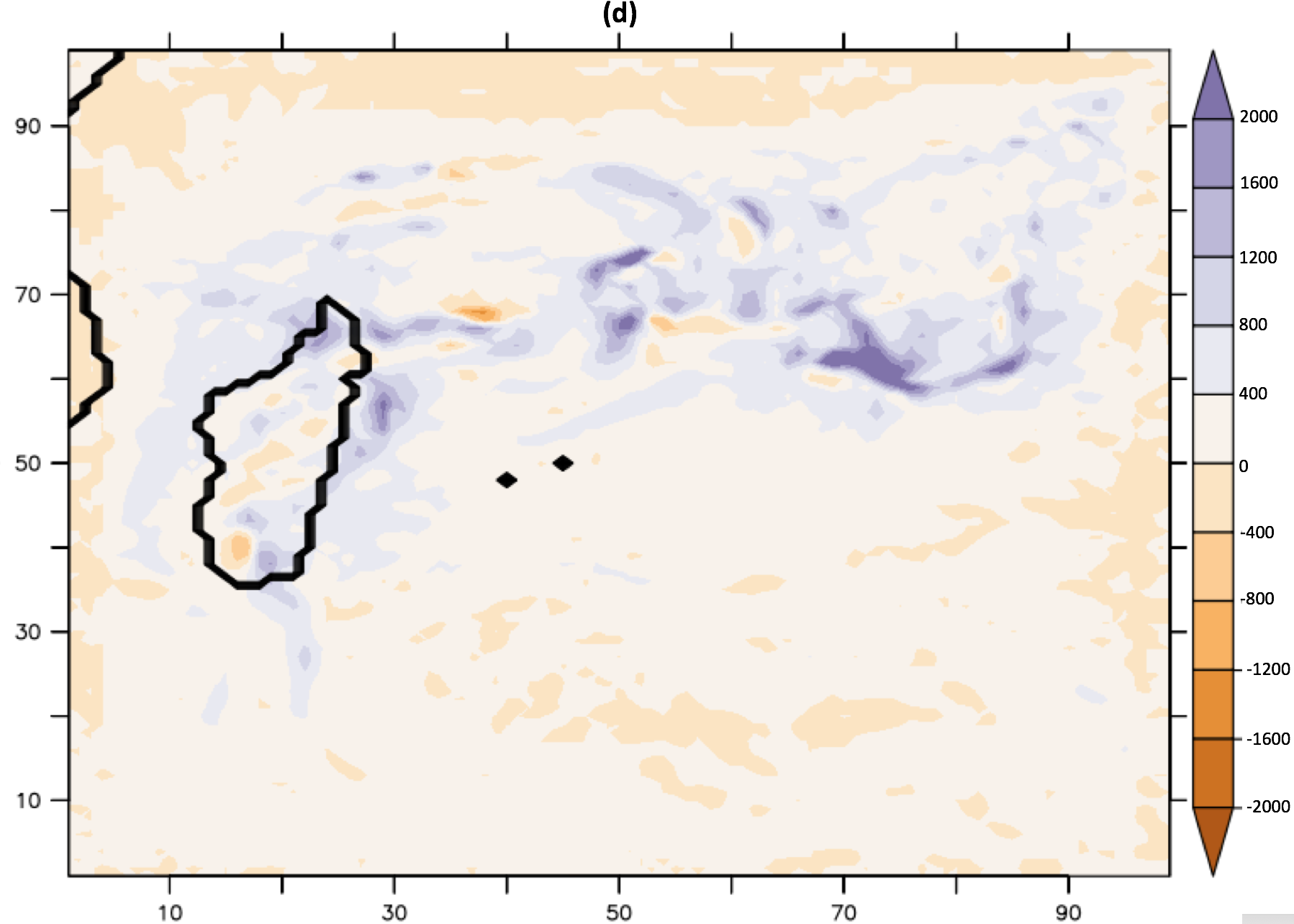}    
    \end{subfigure}%
        \begin{subfigure}{.5\textwidth}
        \centering
\includegraphics[scale=0.22]{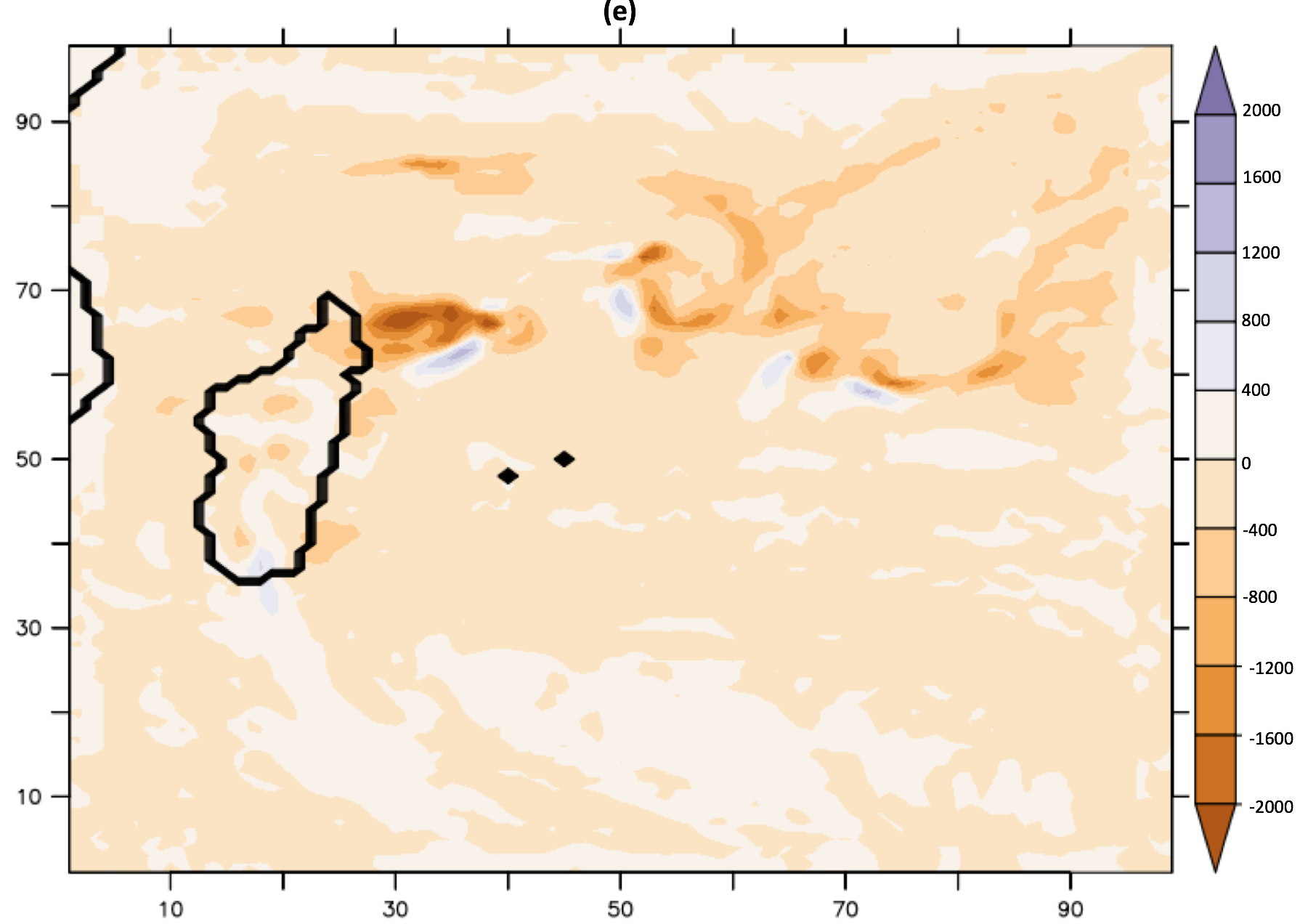}    
    \end{subfigure}%

    \caption{ The total accumulated precipitation on the 6th of March 2017, for the cyclone location in (a) Control simulation, (b) Simulation with SST increased by 2$^{\circ}C$, (c) Simulation with SST increased by 2$^{\circ}C$, (d) Simulation with SST increased by 2$^{\circ}C$ - Control simulation , (e) Simulation with SST decreased by 2$^{\circ}C$ - Control simulation }
    \label{coms}
\end{figure}

It is observed by comparing Figure~\ref{coms} (a) and (b),that an increase in SST causes an increase in the spatial distribution of precipitation pattern over the entire domain. Whereas, a comparison of Figure ~\ref{coms}(a) and (c) illustrates that a decrease in SST leads to a reduction in the spatial distribution of precipitation coverage.
It is also noticed by comparing  Figure~\ref{coms} (a), (b) and (c), that an increase in SST leads to an increase in the cummulative total precipitation over the entire period, while a decrease in SST leads to a decrease.

It is observed from Figure~\ref{coms} (d) that there is a predominant positive spatial value in the difference between the simulation with  SST increased by 2$^{\circ}C$, implying that an increase in SST leads to an overall increase in the total precipitation as well as it's spatial distribution.
On the other hand, from Figure~\ref{coms} (e), there is a predominant zero to negative spatial value in the difference between the simulation with SST decreased by 2$^{\circ}C$. This would imply that a decrease in SST has an effect of decreasing the total precipitation in some areas, but on a larger scale, it has minimal effect on the precipitation.

%\newpage
\subsubsection{Comparative Analysis of Variations in Selected Climatological Variables (Maximum Precipitation Rate, Minimum Surface Pressure and Maximum Windspeed)}
\begin{figure}[ht]
        \begin{subfigure}{.5\textwidth}
            \centering
            \includegraphics[scale=0.28]{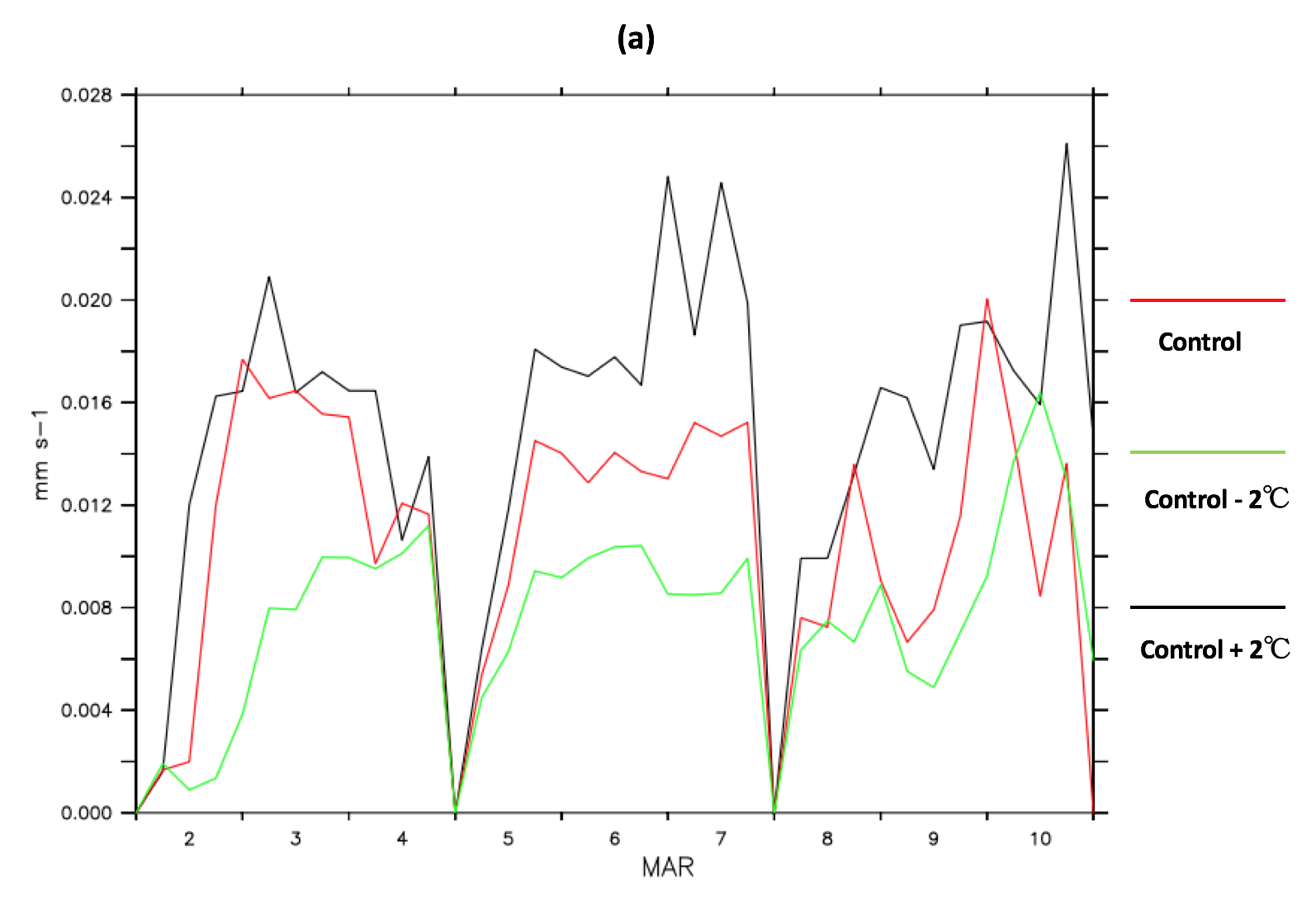}
        \end{subfigure}%
    \begin{subfigure}{.5\textwidth}
        \centering
\includegraphics[scale=0.28]{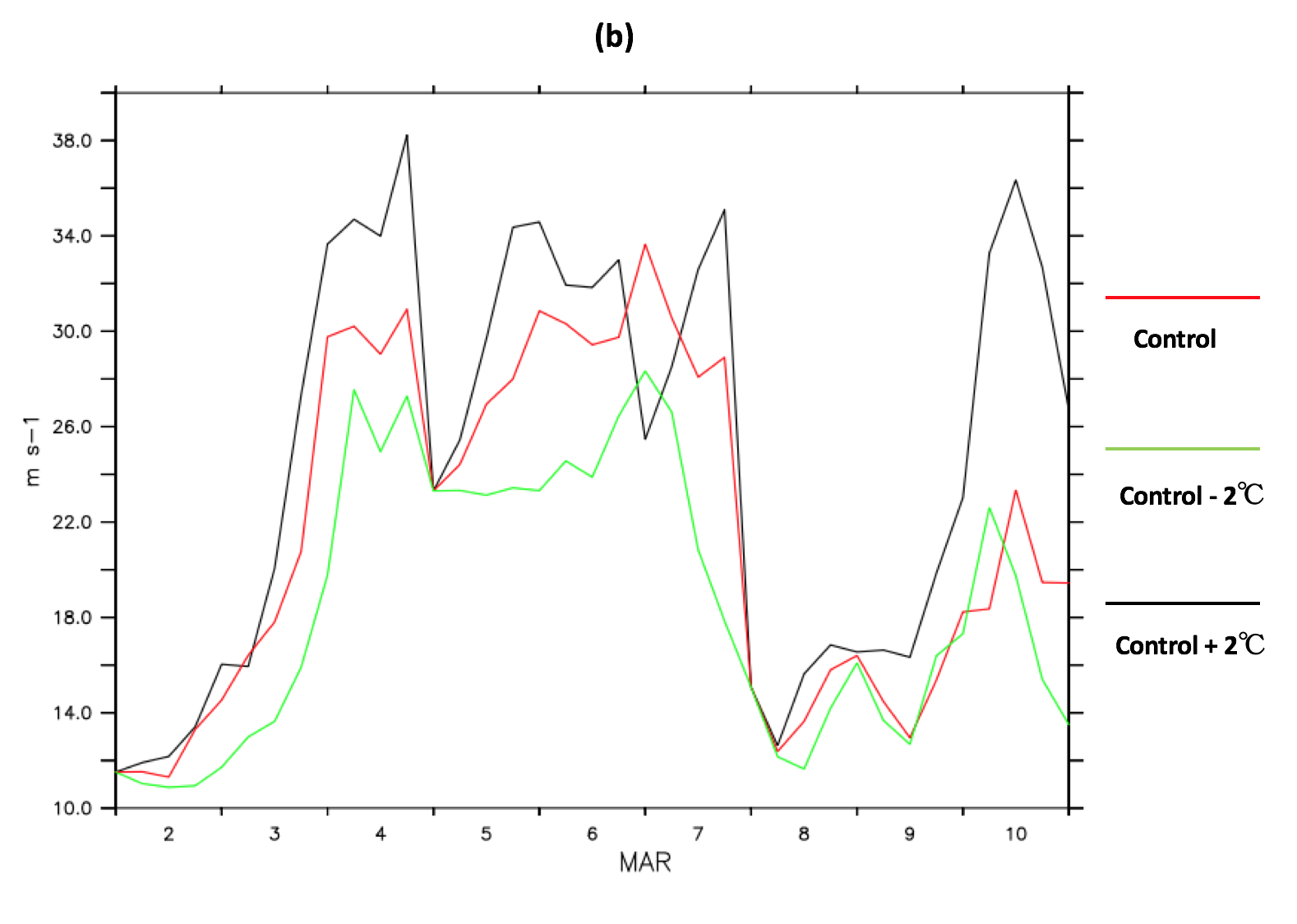}    
    \end{subfigure}%

    \begin{subfigure}{.5\textwidth}
        \centering
\includegraphics[scale=0.28]{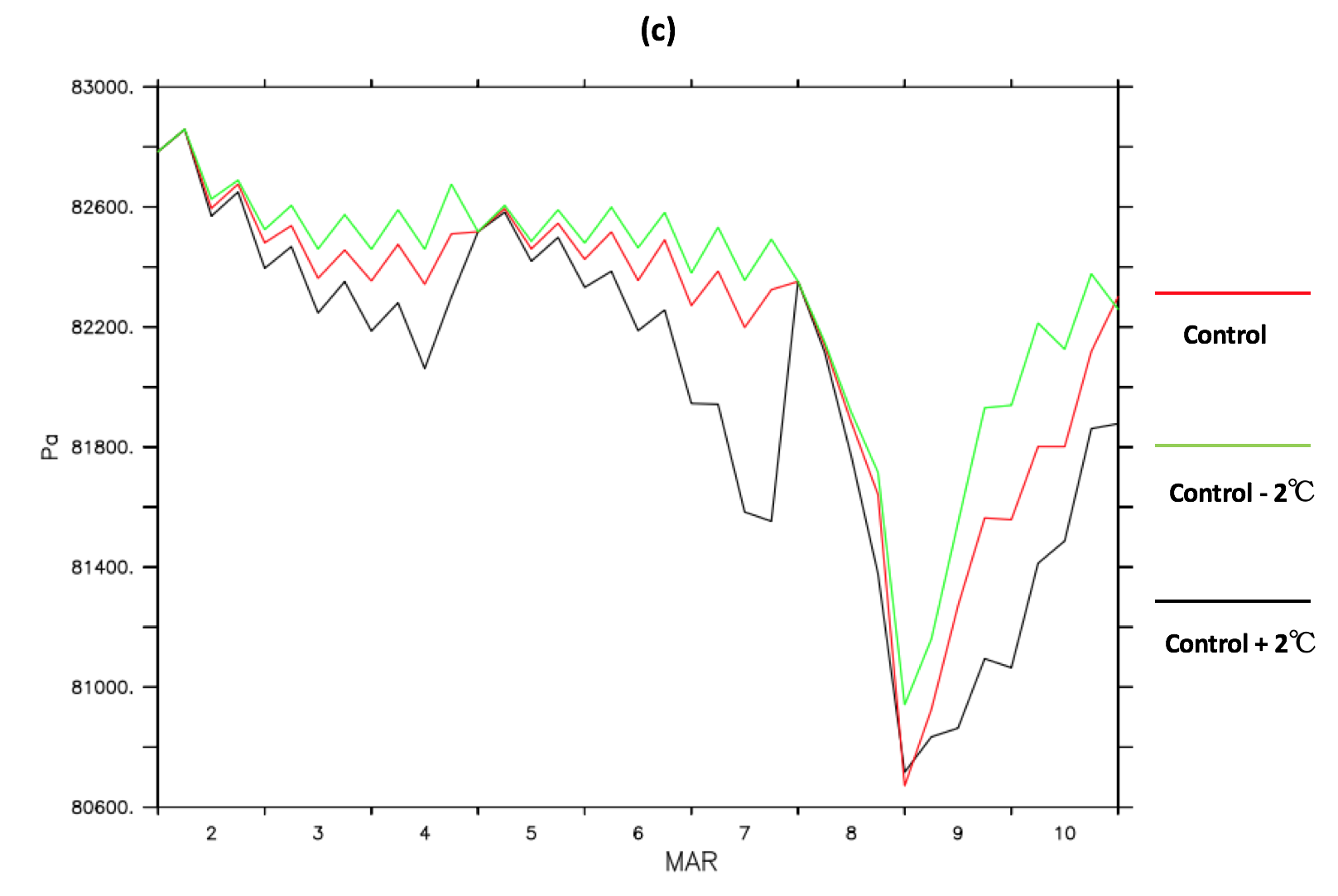}    
    \end{subfigure}%

    \caption{ Figure showing that (a) The maximum precipitation rate for the simulation with increased SST (black) is higher than the control simulation and simulation with decreased sea surface temperature (SST). (b) The maximum value velocity component (\textit{u} at 10m) for the simulation domain is much higher for the simulation with increased sea surface temperature(SST) (black), but lower for the control simulation and simulation with decreased SST. (c) The minimum surface pressure of domain is much lower for the simulation with increased sea surface temperature (SST) (black), and is higher for the control simulation and simulation with decreased SST.}
     \label{joms}
\end{figure}

Figure~\ref{joms} illustrates the  effect of sea surface temperature (SST) on the maximum precipitation rate, minimum surface pressure and maximum windspeed (\textit{velocity at 10m}) for the entire domain within the simulation period which the hurricane lasts (2nd - 11th March 2017).

\qquad Figure~\ref{joms} (a) shows the fluctuation of the maximum precipitation rate over the entire simulation domain  in which the tropical cyclone was tracked by on Figure~\ref{liM}.

We notice the maximum precipitation rate for the control simulation (red line) has a value of about 0.020$mms^{-1}$. This value is less than the maximum precipitation rate for the simulation whose SST was increased by 2$^{\circ}C$ (0.026$mms^{-1}$). This is shown by the black line. Finally, a decrease in the sea surface temperature (SST) of the simulation (shown by green line) illustrates a decrease in the maximum precipitation rate to a value of 0.016$mms^{-1}$.

From these images, we can deduce that an increase in the sea surface temperature (SST) affects the intensity of the tropical cyclone by increasing it's maximum precipitation rate. Also, we can deduce that a decrease in the SST has a negative effect on the tropical cyclone by decreasing it's maximum precipitation rate.

\qquad Figure~\ref{joms} (b) shows a variation in the maximum windspeed (\textit{velocity at 10m}) for the entire domain within the simulation period which the hurricane lasts (2nd - 11th March 2017).

We notice a much more increased value in the vertical windspeed for the simulation with increased SST as opposed to the control simulation and that with decreased SST. For the control simulation (red line), the maximum in windspeed is at about 34$ms^{-1}$. On the other hand, for the simulation with increased SST, there is a maximum windspeed value of 38$ms^{-1}$. Finally, a decrease in the SST produces a maximum windspeed of about 28$ms^{-1}$.

The trend shows that the red-line (simulation with increased SST), is consistently higher than the control simulation (black-line) and the simulation with decreased SST (green-line), illustrating that increasing the SST has an overall effect of increasing the maximum windspeed (\textit{velocity at 10m}) for the entire domain.

\qquad Figure~\ref{joms} (c) shows a variation in the minimum pressure for the entire domain within the simulation period which the hurricane lasts (2nd - 11th March 2017).

We clearly see that an increase in sea surface temperature (shown by the black line) impacts the tropical cyclone by decreasing the minimum surface pressure. However, decreasing the sea surface temperature (SST) has an effect of increasing the minimum pressure measured for the entire domain simulated. This is observed by comparing the green-line and the red-line showing that the green-line is consitently above the black and red lines.
Tropical cyclones are characterized as a low pressure system, thus, the decrease in pressure caused by increasing the SST is indicative of it's positive effect on the intensity of the tropical cyclone.

A combination of the results from Figure~\ref{joms} (a), (b), (c) shows that increasing the sea surface temperature (SST) causes an overall increase in the intensity of the simulated tropical cyclone which manifests in form of increased maximum precipitation rate, decreased surface pressure and increase in the magnitude of the windspeed.

\newpage
\subsubsection{Comparing the Vertical Profiles of the Cyclone Structure}

\begin{figure}[ht]
  \hspace{1.63in}
          \begin{subfigure}{.5\textwidth}
        \centering
\includegraphics[scale=0.22]{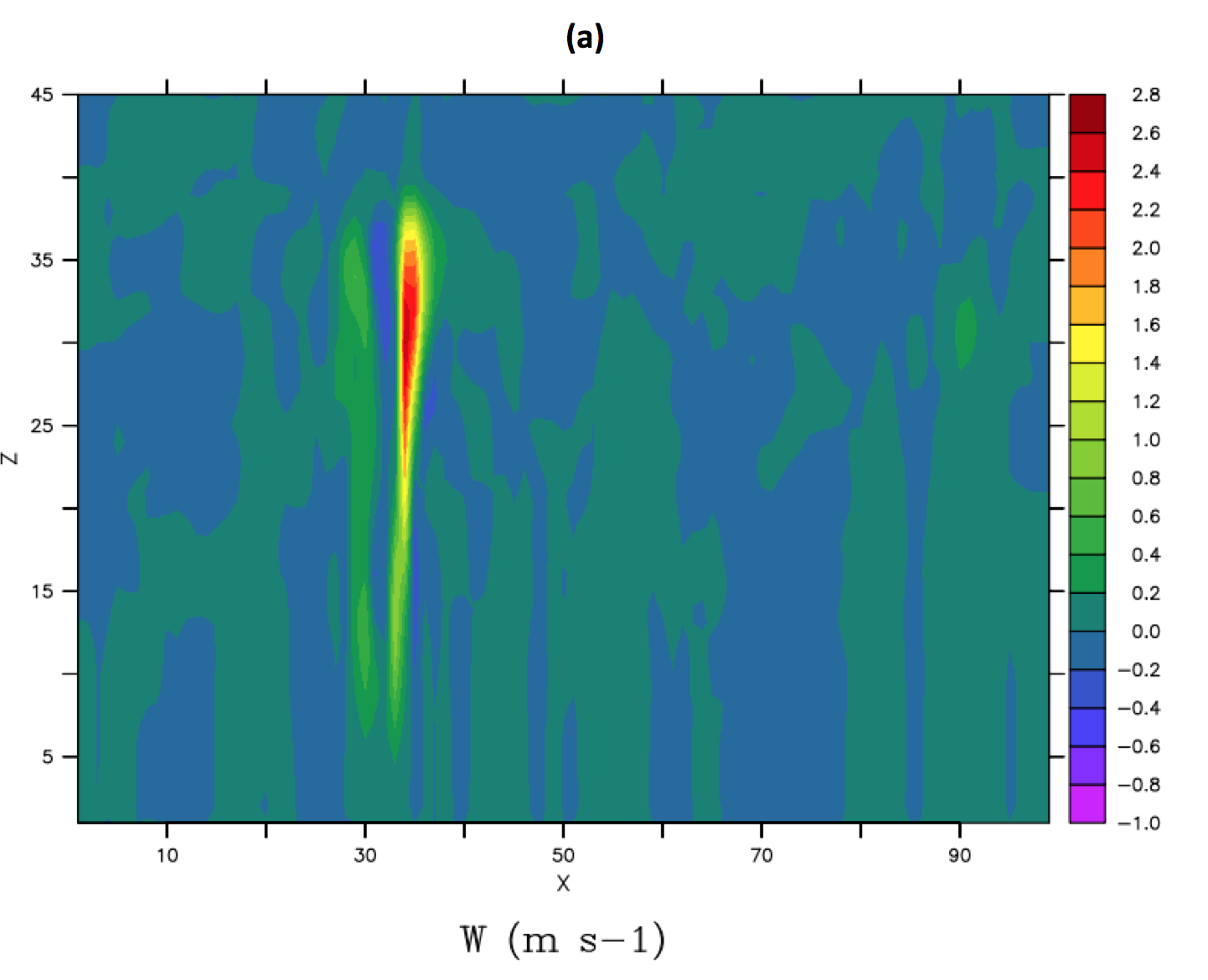}    
    \end{subfigure}%
    
        \begin{subfigure}{.5\textwidth}
            \centering
            \includegraphics[scale=0.22]{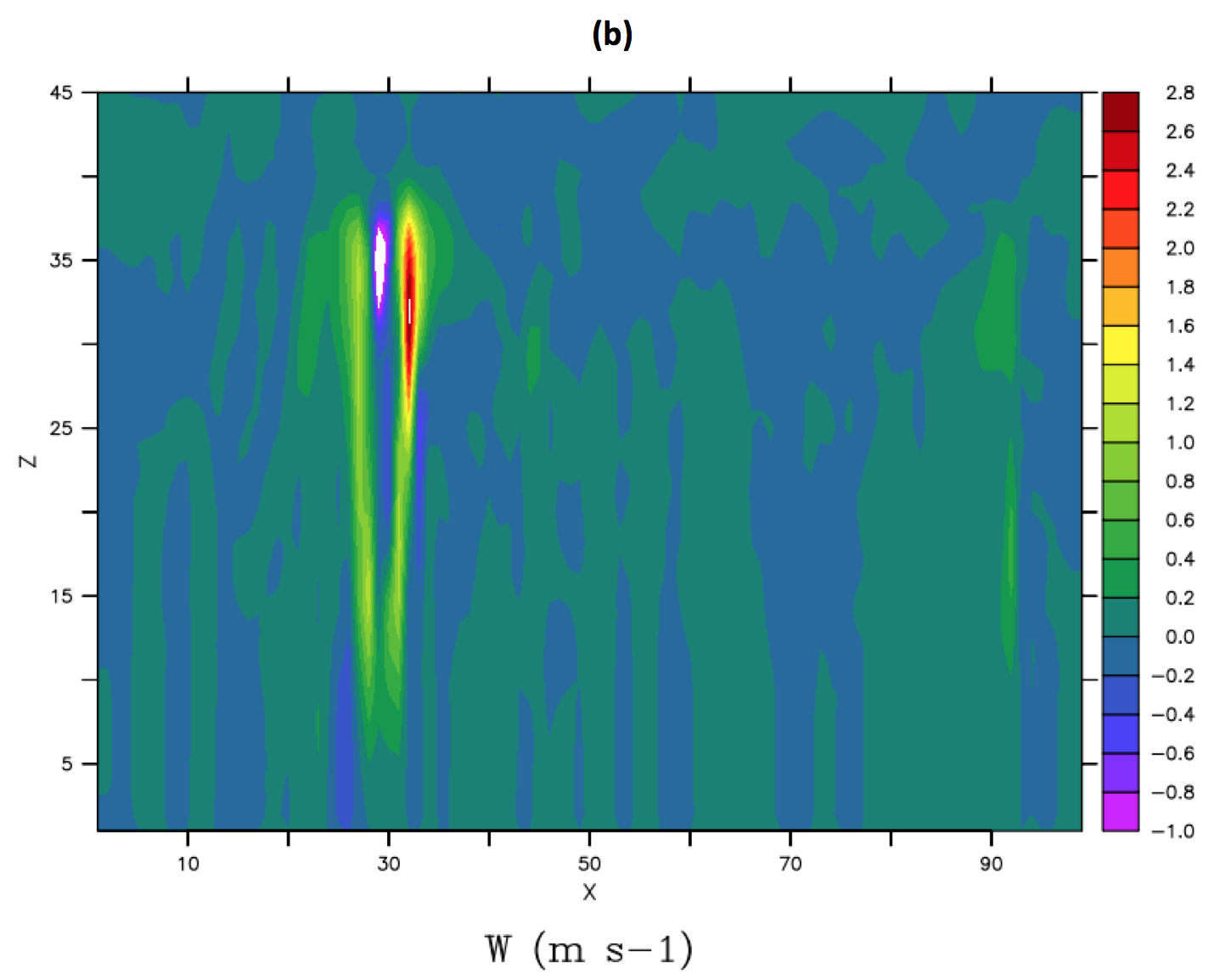}
        \end{subfigure}%
    \begin{subfigure}{.5\textwidth}
        \centering
\includegraphics[scale=0.22]{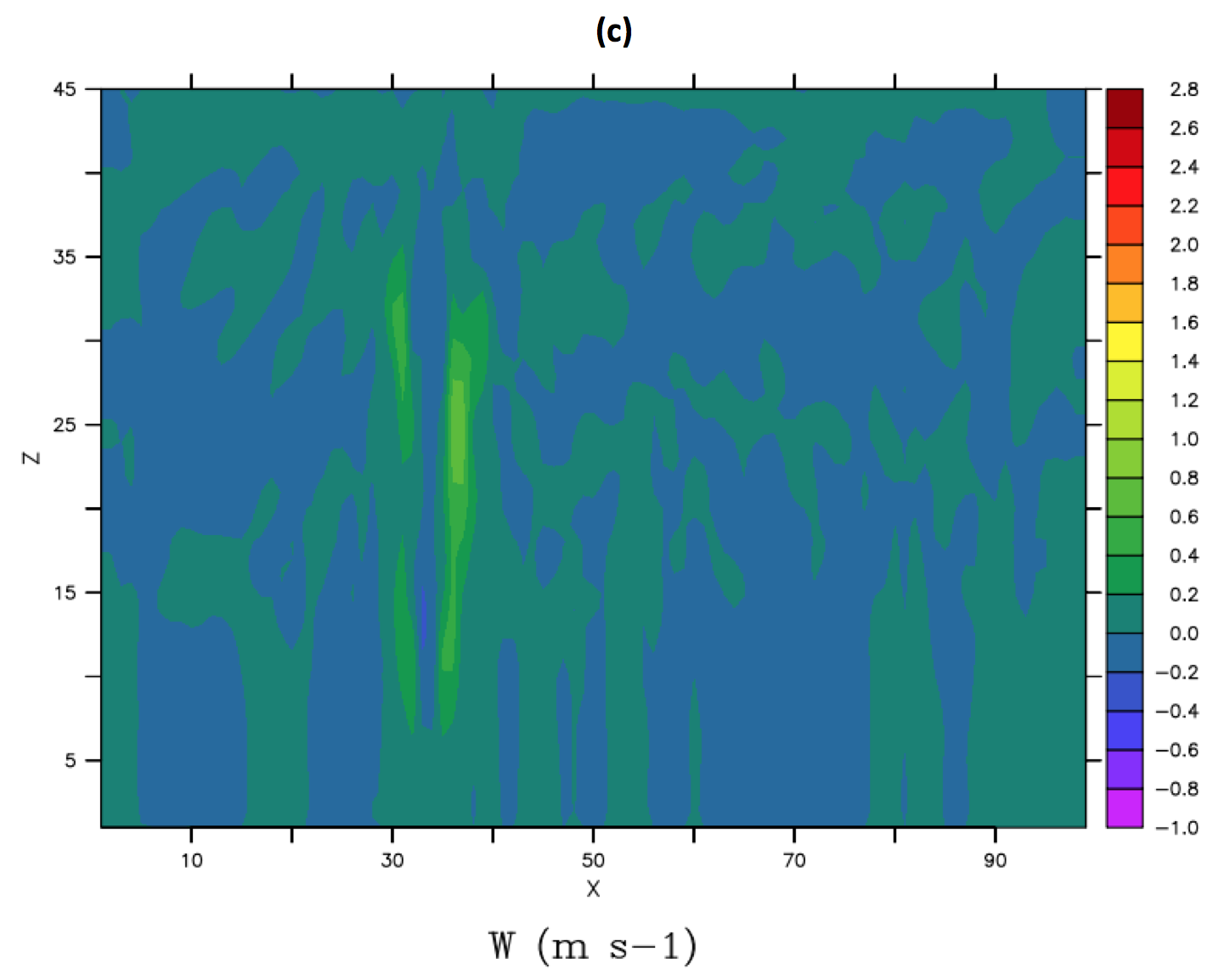}    
    \end{subfigure}%

    \begin{subfigure}{.5\textwidth}
        \centering
\includegraphics[scale=0.24]{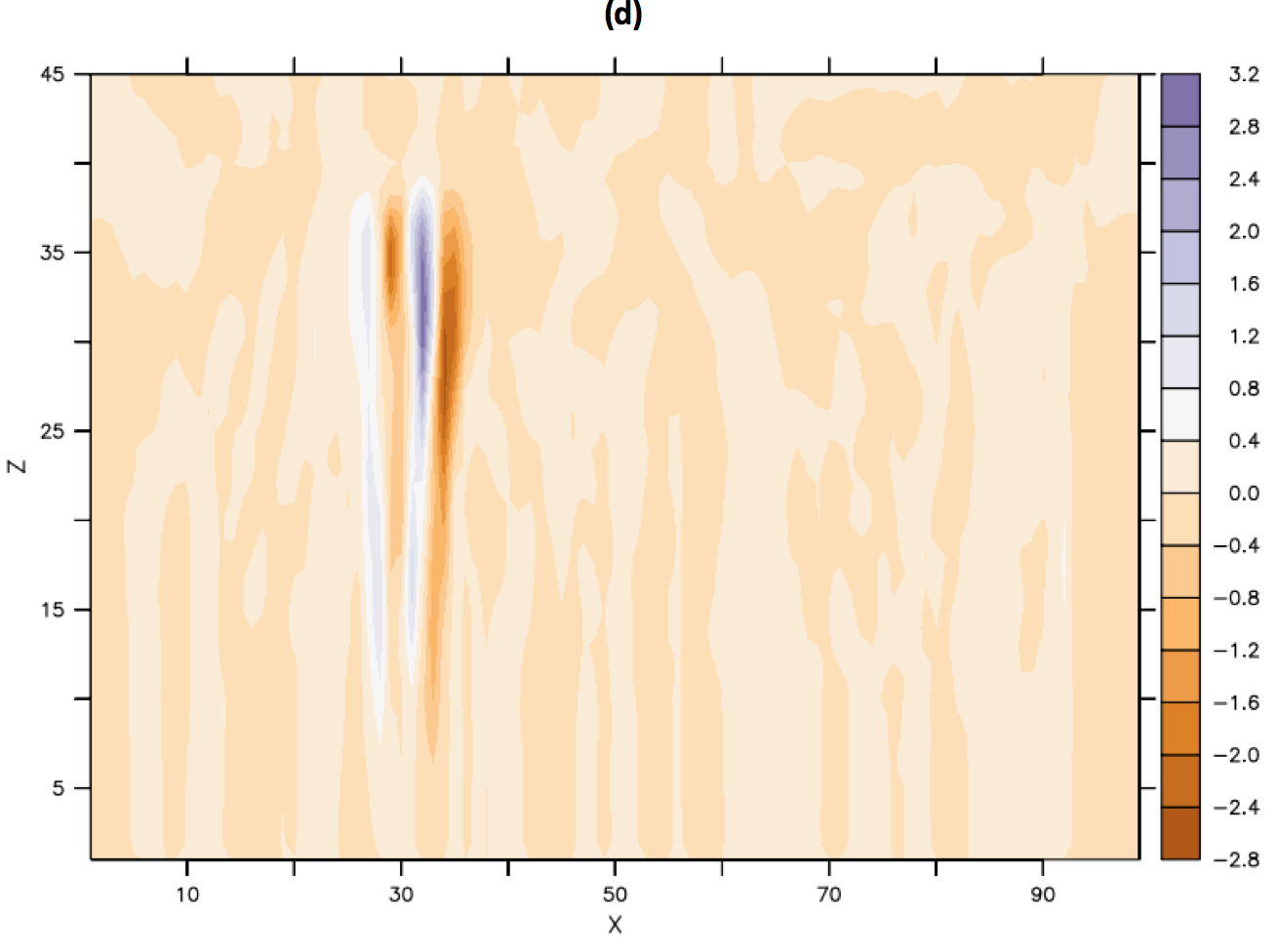}    
    \end{subfigure}%
        \begin{subfigure}{.5\textwidth}
        \centering
\includegraphics[scale=0.24]{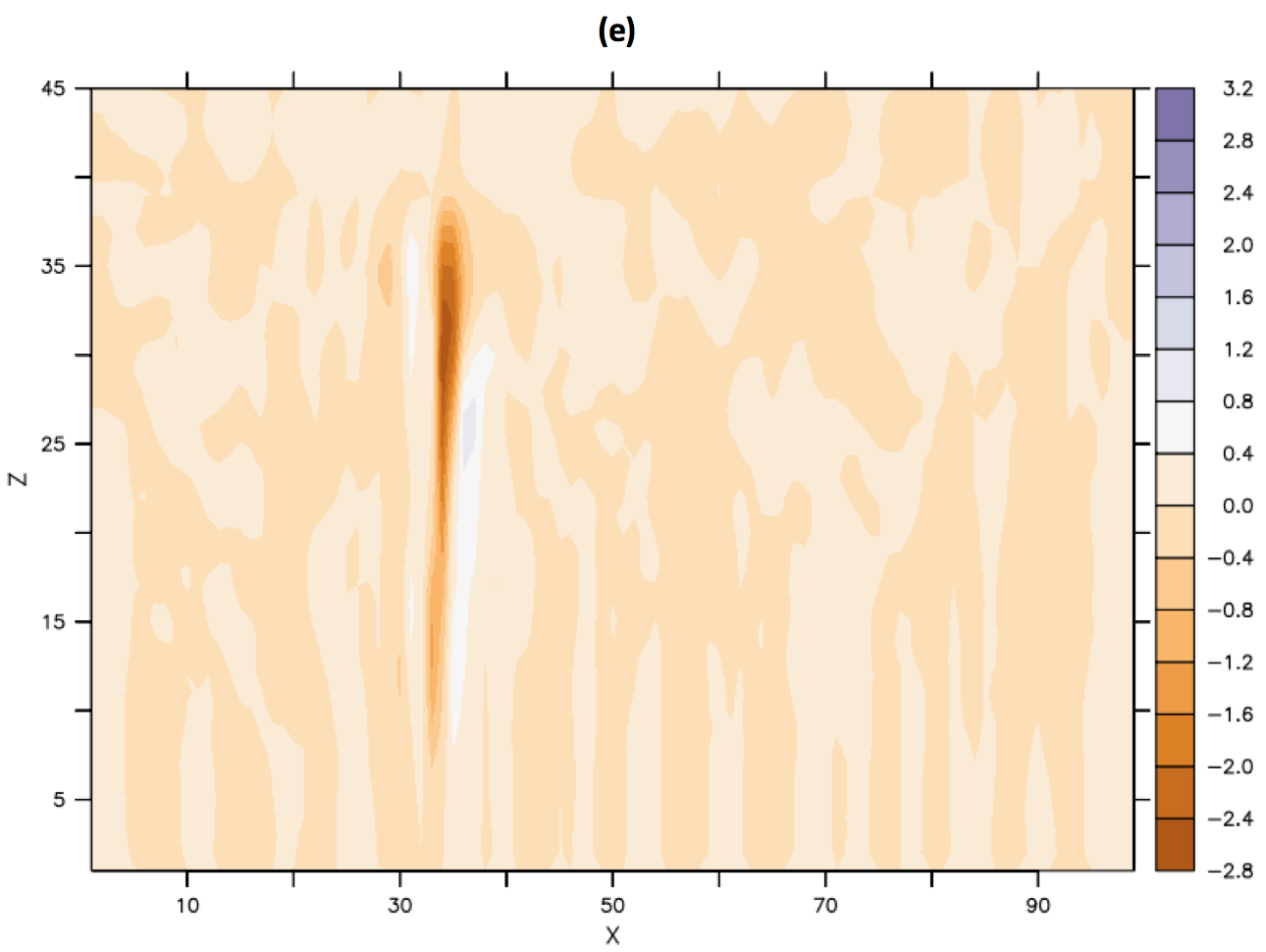}    
    \end{subfigure}%

\caption{ Vertical profiles showing variation in vertical windspeed of the tropical cyclone for: (a) The control simulation (b) The simulation with increased  sea surface temperature (SST) by 2$^{\circ}C$, (c) The simulation with decreased sea surface temperature (SST) by 2$^{\circ}C$, (d) Simulation with SST increased by 2$^{\circ}C$ - Control simulation, (e) Simulation with SST decreased by 2$^{\circ}C$ - Control simulation }
\label{ed}
\end{figure}

\begin{figure}[ht]
  \hspace{1.63in}
          \begin{subfigure}{.5\textwidth}
        \centering
\includegraphics[scale=0.21]{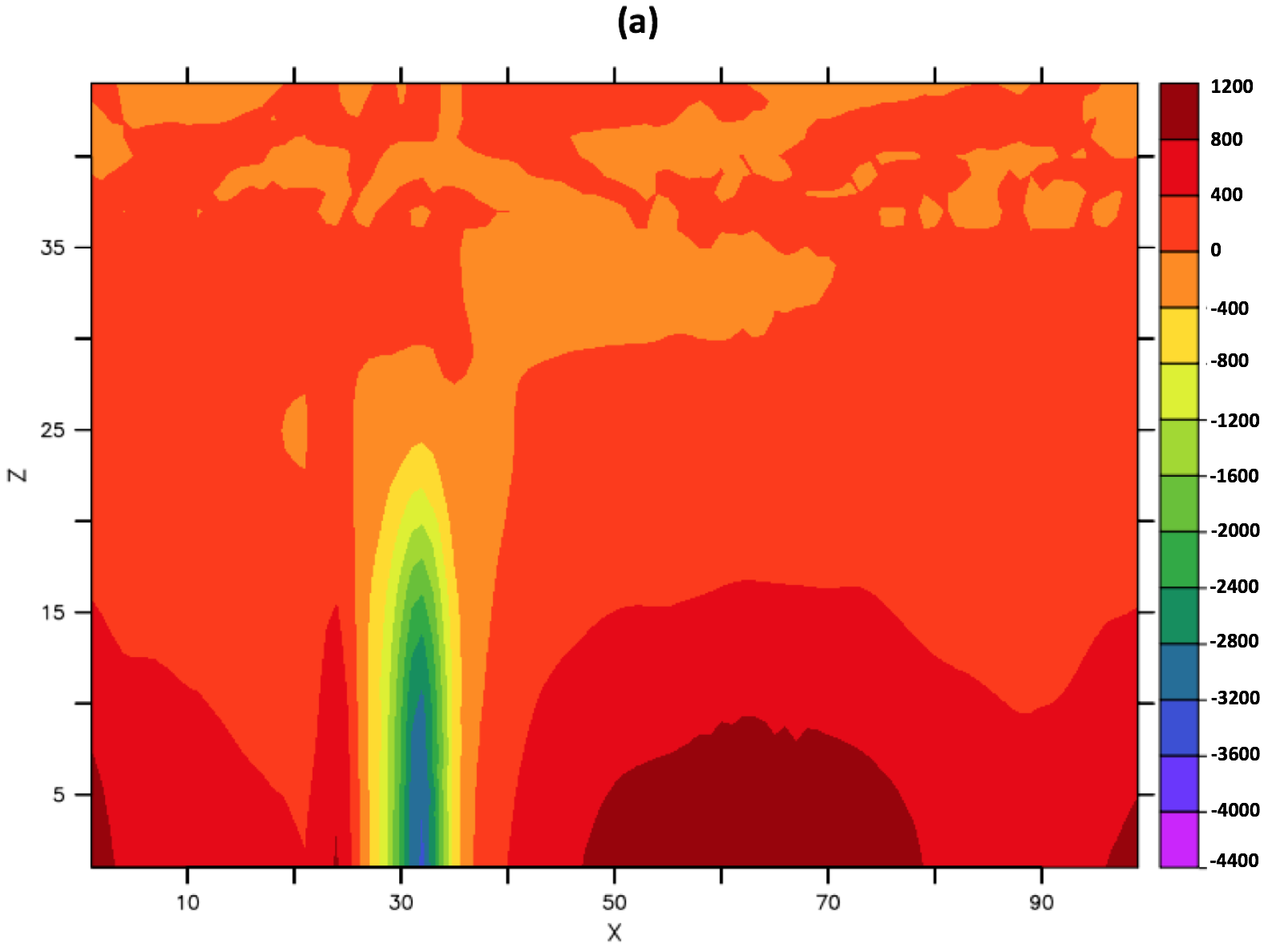}    
    \end{subfigure}%
    
        \begin{subfigure}{.5\textwidth}
            \centering
            \includegraphics[scale=0.21]{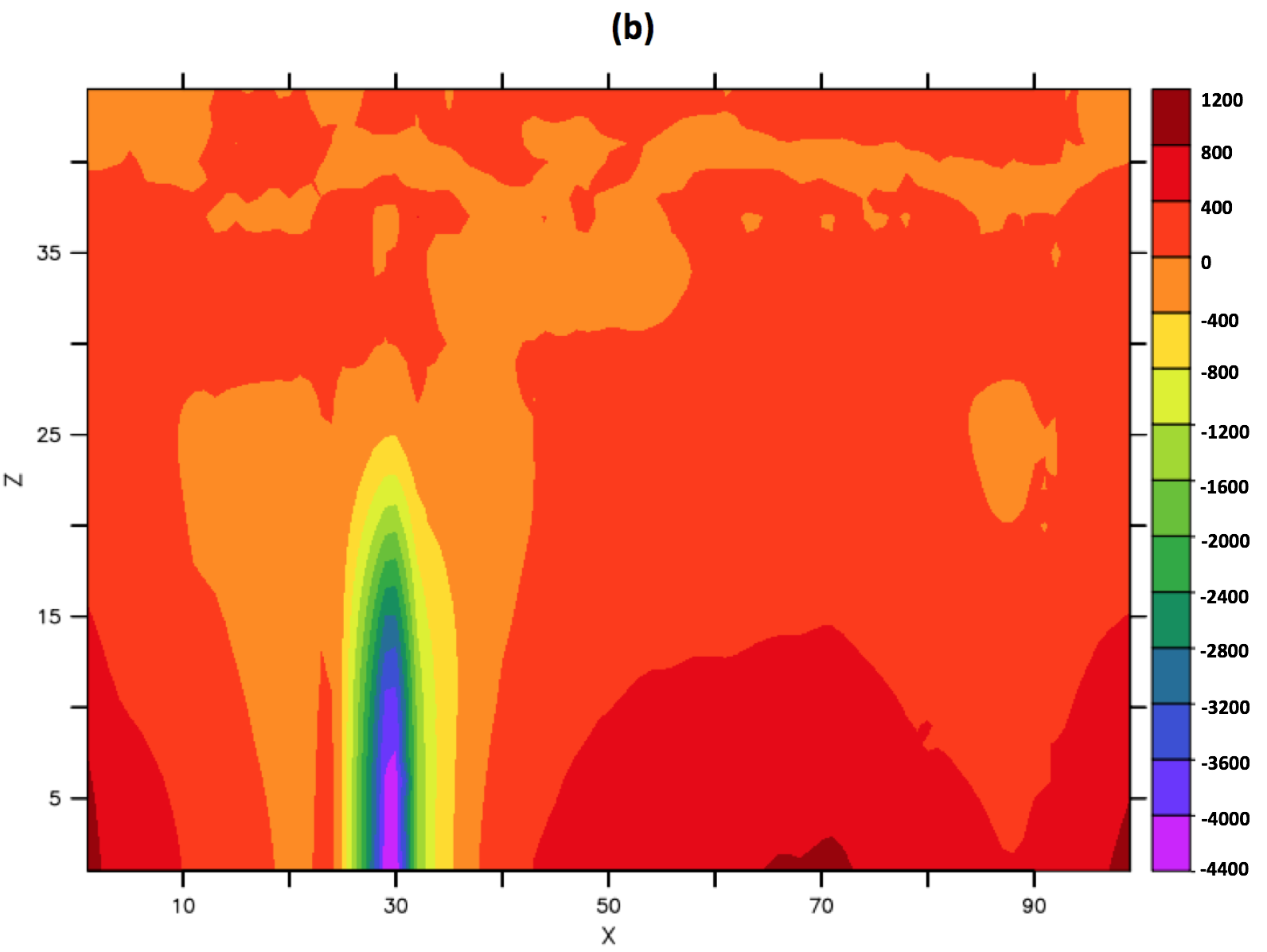}
        \end{subfigure}%
    \begin{subfigure}{.5\textwidth}
        \centering
\includegraphics[scale=0.21]{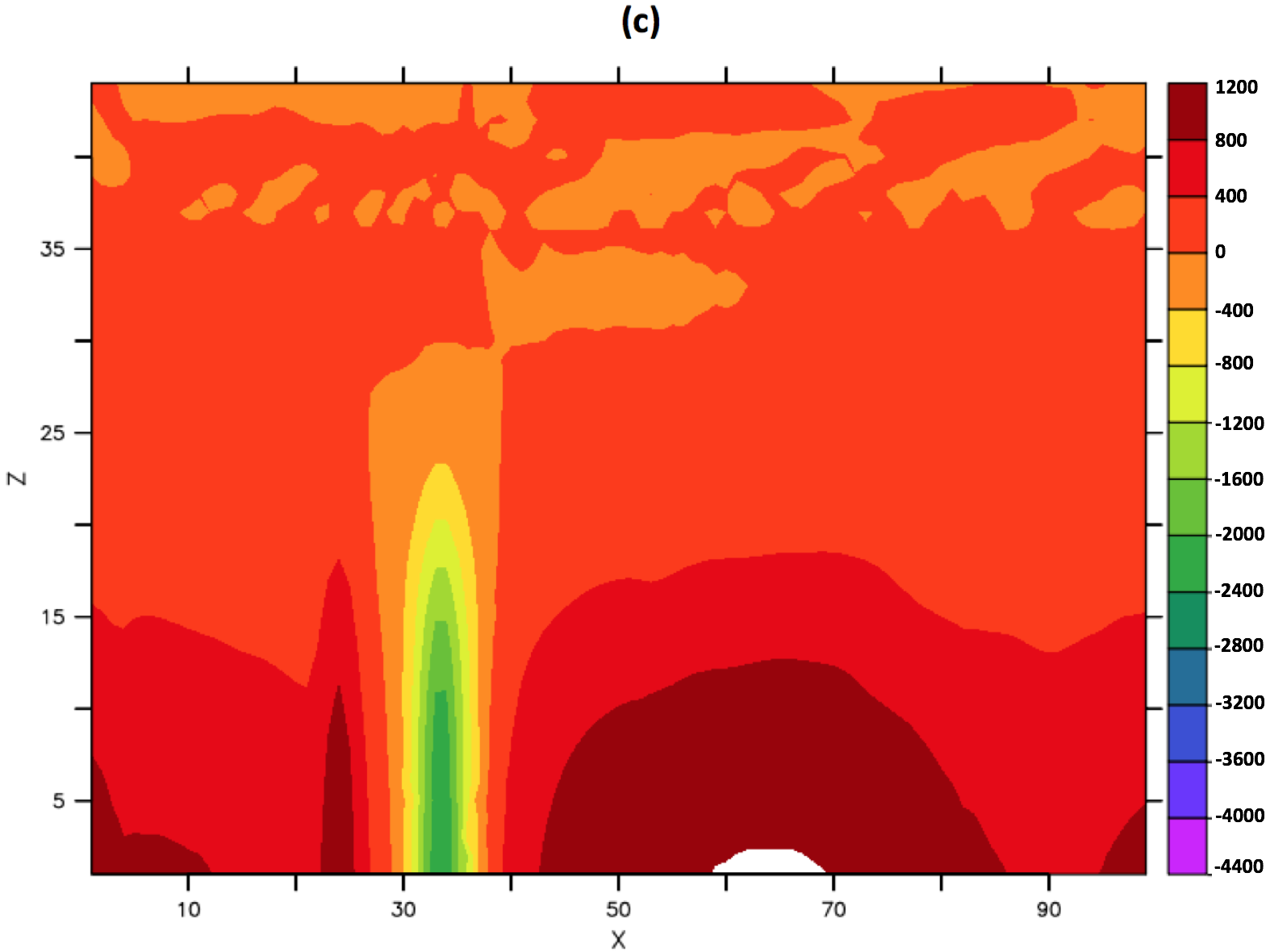}    
    \end{subfigure}%

    \begin{subfigure}{.5\textwidth}
        \centering
\includegraphics[scale=0.20]{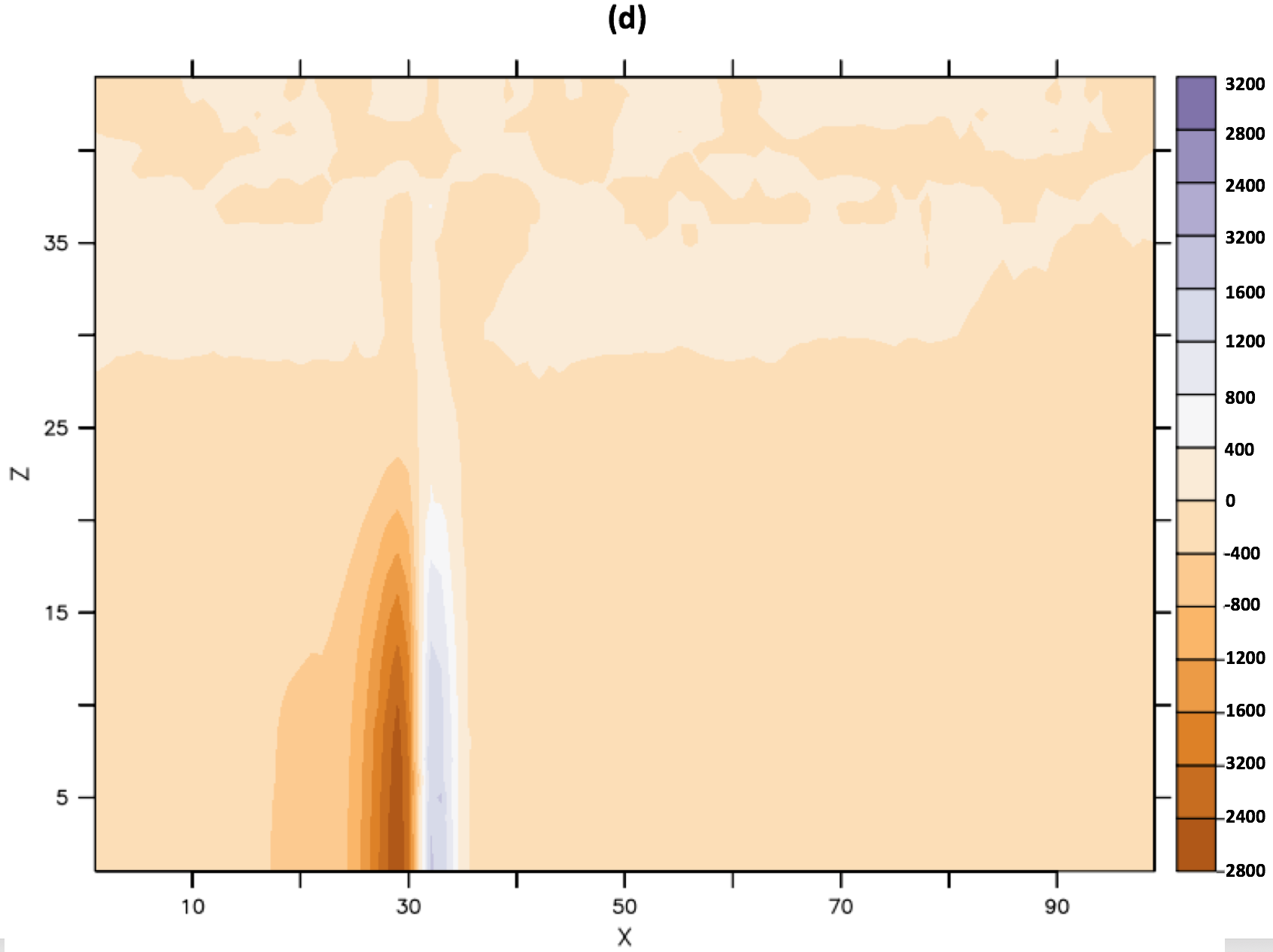}    
    \end{subfigure}%
        \begin{subfigure}{.5\textwidth}
        \centering
\includegraphics[scale=0.20]{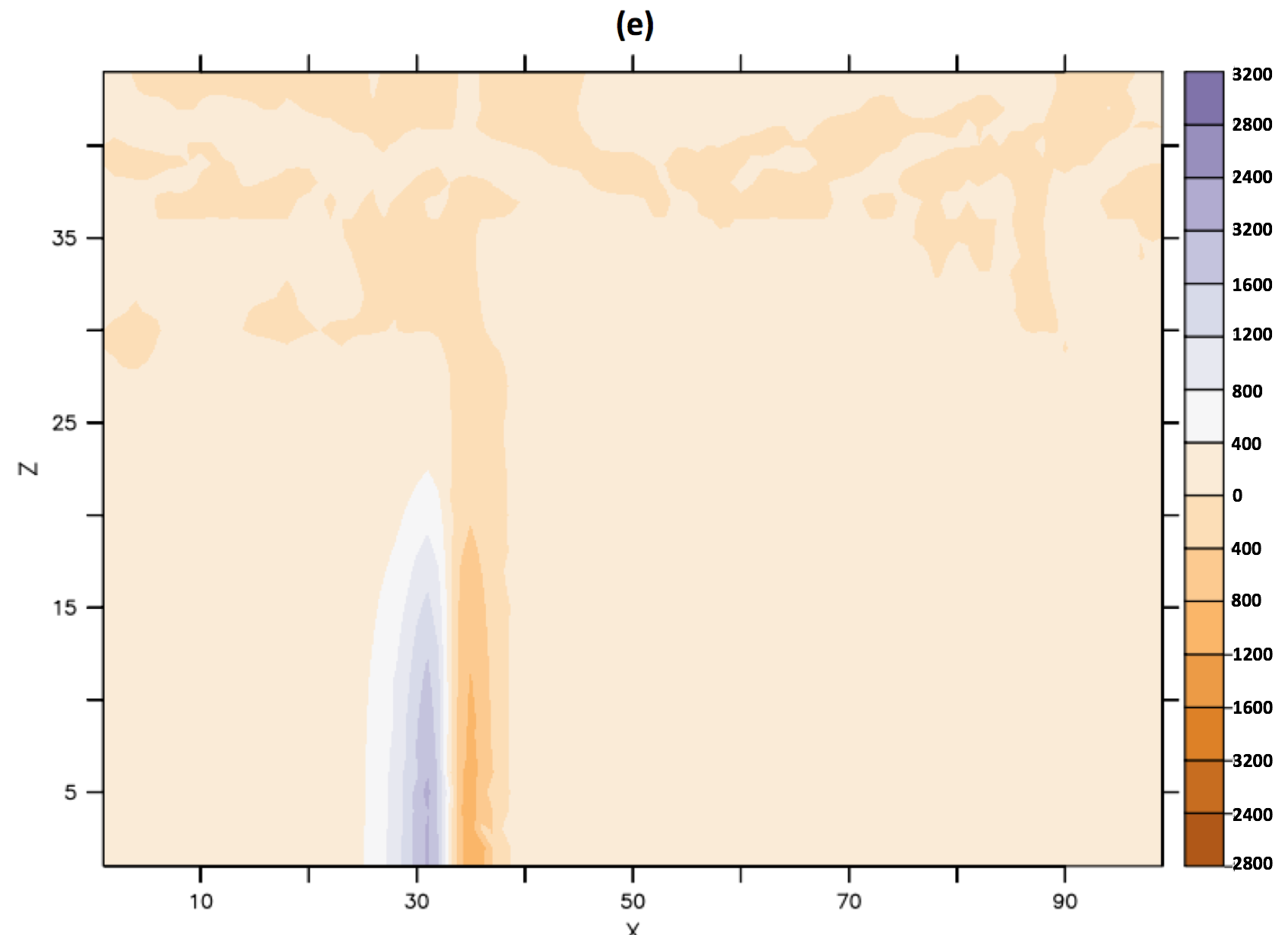}    
    \end{subfigure}%

\caption{Vertical profiles showing variation in pressure of the tropical cyclone for: (a.) The control simulation, (b) The simulation with increased  sea surface temperature (SST) by 2$^{\circ}C$, (c) The simulation with decreased  sea surface temperature (SST) by 2$^{\circ}C$, (d) Simulation with SST increased by 2$^{\circ}C$ - Control simulatio , (e) Simulation with SST decreased by 2$^{\circ}C$ - Control simulation}
\label{eddy}
\end{figure}

Figures~\ref{ed} (a) - (e) and Figures~\ref{eddy} (a) - (e) shows a vertical profile of the windspeed and pressure for the tropical cyclone indicating variations in intensity for the control simulation, simulation with increased SST and simulation with decreased SST.

A comparison of Figures~\ref{ed} (a) and (b) shows that there is a decrease in the windspeed at the eye of the tropical cyclone (30$^{\circ}$) for the simulation with SST increased by 2$^{\circ}C$. Also, Figures~\ref{ed} (d) shows a negative difference at the eye of the tropical cyclone, also indicative of a decreasing windspeed with increased SST.

On the other hand, by comparing Figures~\ref{ed} (a) and (c), there is no major difference in the magnitude of windspeed at the eye. However, there is a decrease in the magnitude of the windspeed at the eye wall. This is indicative of a decreasing strenght in the strenght of the tropical cyclone. Also, Figures~\ref{ed} (e) shows a zero to positive difference at the eye. This indicates that a decrease in SST has little or no effect on the intensity of the this tropical cylone.

\newpage
A comparison of Figures~\ref{eddy}(a) and (b) shows that there is a decrease in the pressure at the core of the tropical cyclone for the simulation with increased SST. Figures~\ref{eddy} (d) shows a negative difference at the cylone core,indicative of a decreasing core pressure with increased SST.

Whereas, Figures~\ref{eddy} (a) and (c), there is barely any noticeable difference in the core pressure. Figures~\ref{eddy} (e) shows a zero to minute positive difference at the eye. This indicates that a decrease in SST has little effect on the intensity of the tropical cyclone, acting to increase the core pressure.

\newpage
\subsubsection{Sensitivity of Windspeed Vector}
\begin{figure}[ht]
        \begin{subfigure}{.5\textwidth}
            \centering
            \includegraphics[scale=0.28]{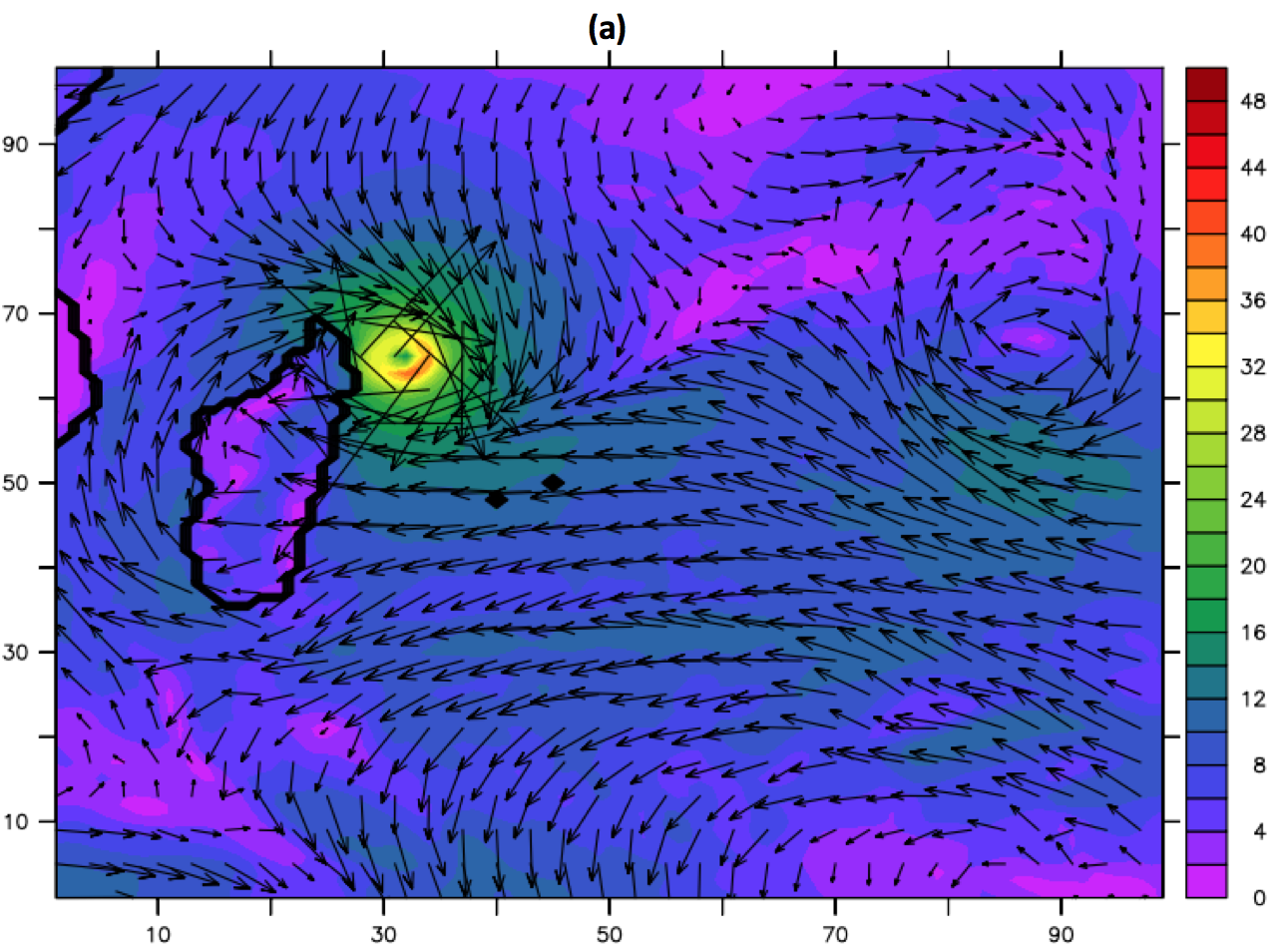}
        \end{subfigure}%
    \begin{subfigure}{.5\textwidth}
        \centering
\includegraphics[scale=0.28]{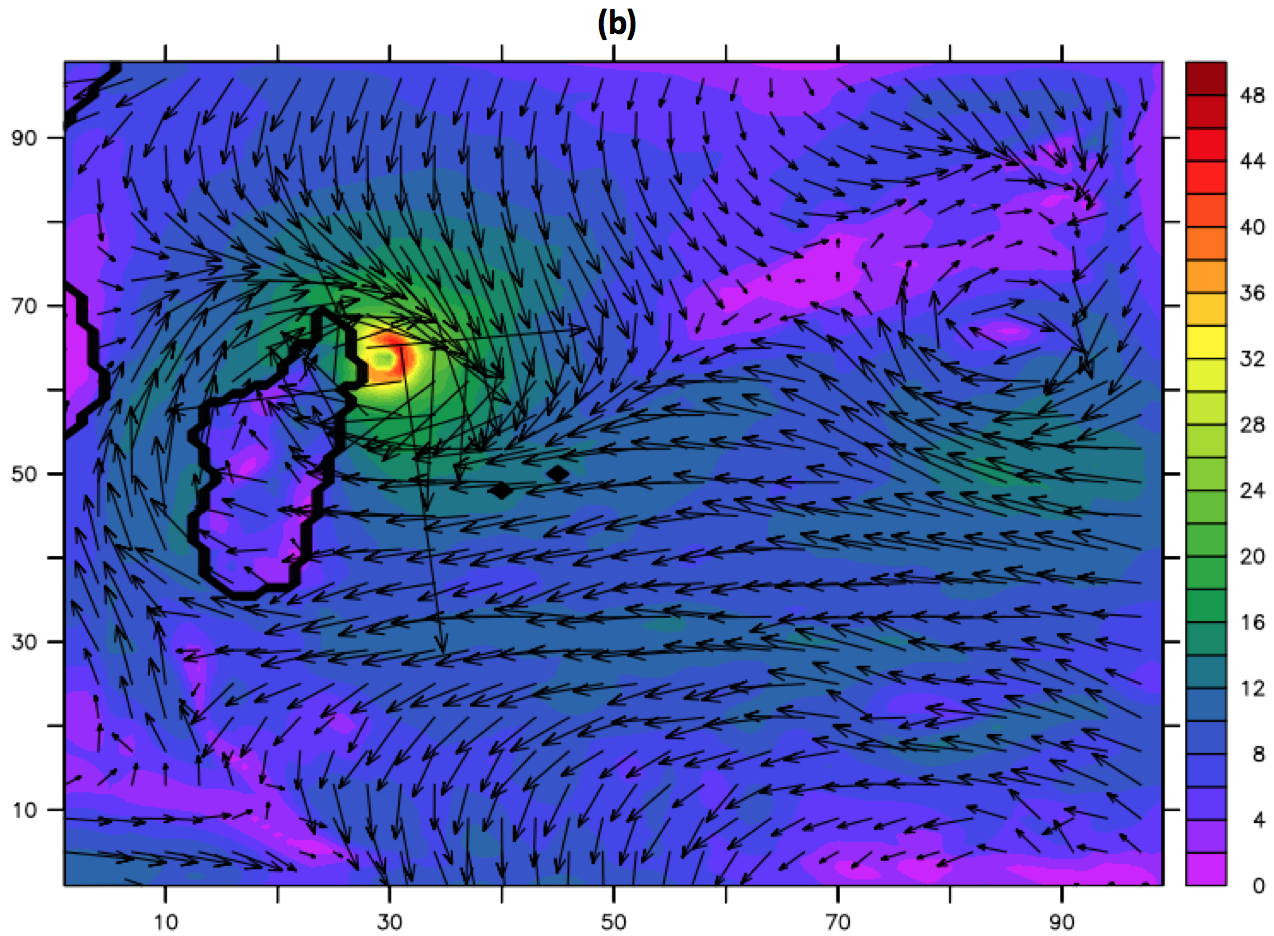}    
    \end{subfigure}%

    \begin{subfigure}{.5\textwidth}
        \centering
\includegraphics[scale=0.28]{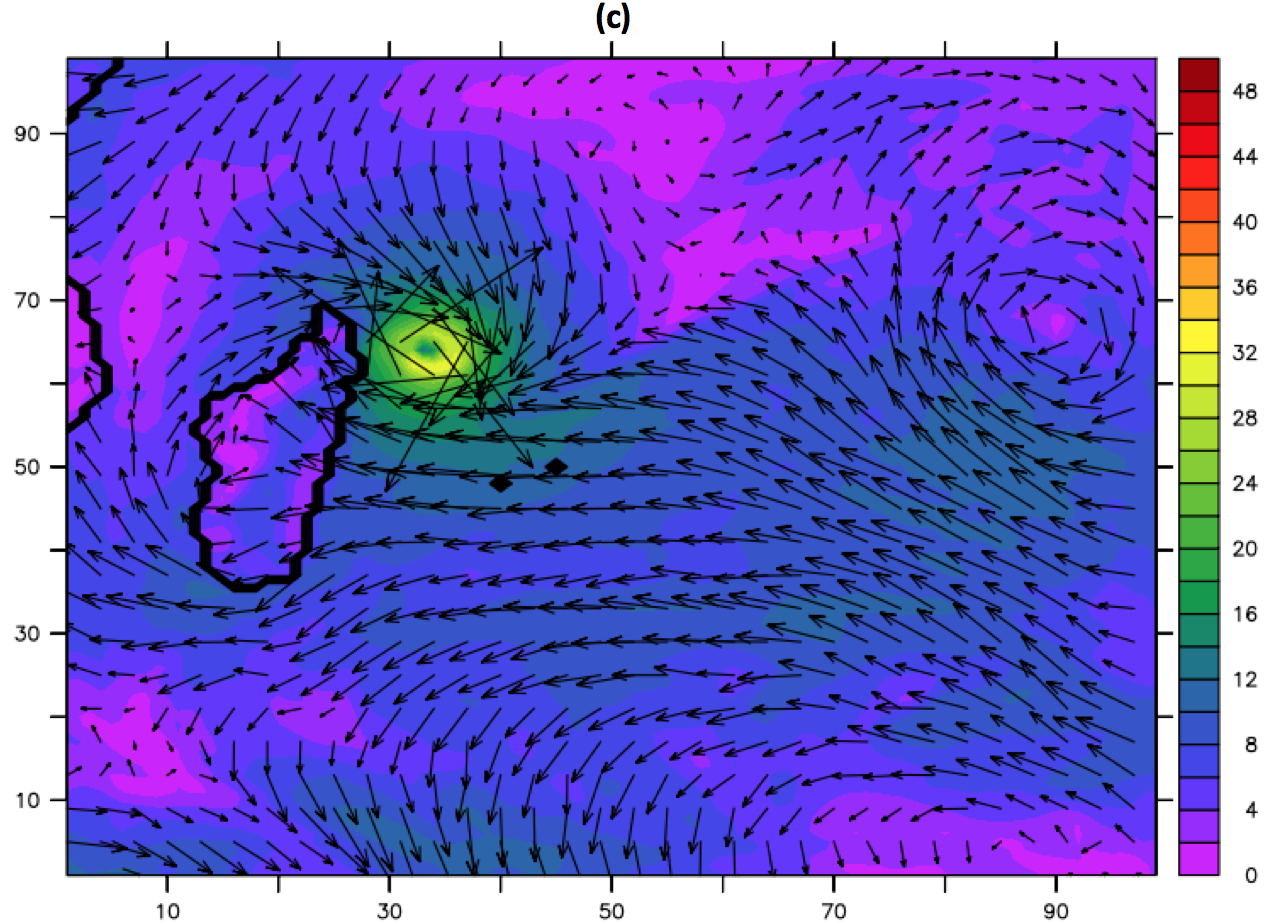}    
    \end{subfigure}%

    \caption{(a) The velocity vector field for the control simulation (b) the velocity vector field for  simulation with increased  sea surface temperature (SST) (c) the velocity vector field for  simulation with decreased  sea surface temperature (SST).}
     \label{fight}
\end{figure}

Figure~\ref{fight} shows the vector of windspeed component on the 6th of March for the location in which the tropical cyclone was tracked by Figure~\ref{liM} to be located. 

Figure~\ref{fight} (a) indicates maximum windspeed component of about 38$ms^{-1}$ for the control simulation. However, Figure~\ref{fight} (b) shows maximum windspeed component of about 46$ms^{-1}$ for the simulation with increased  sea surface temperature (SST) of 2$^{\circ}C$, while Figure~\ref{fight} (c) shows maximum windspeed component of about 30$ms^{-1}$ for the simulation with decreased  sea surface temperature (SST) of 2$^{\circ}C$.

The wind vector also shows a much more stronger curl vector field in Figure~\ref{fight} (b), as compared with Figure~\ref{fight} (a) and Figure~\ref{fight} (c), indicating that an increase in sea surface temperature (SST) affects the intensity of tropical cyclone by increasing it's windspeed.

\section{Conclusion and Recommendation}

\subsection{Summary}
In a bid to understand the possible effect that global warming and a violation of the Paris accord could have on tropical cyclogenesis and it's intensification around the South West Indian ocean, this study investigated the influence of sea surface temperature (SST) variations in simulated tropical cyclones from the WRF regional climate model. The objective of this study was to observe the sensitivity of tropical cyclone Enawo to increase of 2$^{\circ}C$ in sea surface temperature (SST), as well as  decrease of 2$^{\circ}C$. The characteristics of the tropical cyclone Enawo which was observed for sensitivity analysis include the cyclone track, precipitation, surface pressure, wind speed vector and the vertical cross-section/struture of the tropical cyclone.

In order to achieve these objectives, the simulations from the WRF model was initially validated using the European Centre for Medium-Range Weather Forecasts (ECMWF) ERA5 re-analysis dataset for the active period of tropical cyclone Enawo (2nd - 11th March 2017). The effectiveness of the WRF model in representing the actual properties of the tropical cyclone was obtained by comparing it's output with that obtained from European Centre for Medium-Range Weather Forecasts (ECMWF) ERA5 re-analysis dataset . To investigate the role of sea surface temperature (SST) on the WRF simulations of the cyclone, the SST from the boundary conditions for the WRF model was increased by 2$^{\circ}C$ before simulation. The same procedure was repeated but with a decrease in SST of 2$^{\circ}C$ as well.

The result of the study is summarized below:

\begin{itemize}
\item The simulation of cyclone Enawo using the WRF model is sensitive to increase/decrease in the sea surface temperature (SST) of the boundary dataset used in the simulation. However, it is more sensitive to temperature increase than a decrease.

\item Increasing the SST by 2$^{\circ}C$ has an overall increase in the intensity of the tropical cyclone. This increase is manifested as:
	\begin{itemize}
		\item Increase in the maximum precipitation rate.
		\item Increase in the magnitude of the maximum windspeed.
		\item Decrease in the magnitude of the minimum surface pressure.
	\end{itemize}
	
\item Decreasing the SST by 2$^{\circ}C$ has an overall decrease in the intensity of the tropical cyclone. This increase is manifested as:
	\begin{itemize}
		\item Decrease in the maximum precipitation rate.
		\item Decrease in the magnitude of the maximum windspeed.
		\item Increase in the magnitude of the minimum surface pressure.
	\end{itemize}
 
\item Decreasing the SST by 2$^{\circ}C$ has a less pronounce effect on the track followed by tropical cyclone Enawo. The decrease has an overall effect of smoothening the track followed by the cyclone. On the other hand, increasing the SST by 2$^{\circ}C$ has a more cospicuous effect on the track followed by cyclone Enawo. This is evident in the emergence of a second low-pressure system on the 10th of March. This indicates that an increase in SST leads to an increase in the number of occurences of tropical systems.

\end{itemize}

\subsection{Further work}
The robustness of results obtained from this study can be improved by providing multiple simulations of tropical cyclone Enawo using different regional climate models. This would help to reinforce the confirmation of the sensitivity of tropical cyclones to sea surface temperature (SST). Further more, while the current study focusses on a single resolution, we could observe the effect of various model resolution on the results obtained.

\newpage
\bibliographystyle{unsrt}  
\bibliography{references}  %%% Remove comment to use the external .bib file (using bibtex).

%%%\begin{thebibliography}{1}

%%%\bibitem{kour2014real}
%%%George Kour and Raid Saabne.
%%%\newblock Real-time segmentation of on-line handwritten arabic script.
%%%\newblock In {\em Frontiers in Handwriting Recognition (ICFHR), 2014 14th
 %%% International Conference on}, pages 417--422. IEEE, 2014.

%%%\bibitem{kour2014fast}
%%%George Kour and Raid Saabne.
%%%\newblock Fast classification of handwritten on-line arabic characters.
%%%\newblock In {\em Soft Computing and Pattern Recognition (SoCPaR), 2014 6th
 %%% International Conference of}, pages 312--318. IEEE, 2014.

%%%\bibitem{hadash2018estimate}
%%%Guy Hadash, Einat Kermany, Boaz Carmeli, Ofer Lavi, George Kour, and Alon
 %%% Jacovi.
%%%\newblock Estimate and replace: A novel approach to integrating deep neural
 %%% networks with existing applications.
%%%\newblock {\em arXiv preprint arXiv:1804.09028}, 2018.

%%%\end{thebibliography}

\end{document}